\documentclass[prd,twocolumn,nopacs,floatfix,amsmath,nofootinbib,amssymb,floatfix]{revtex4}
\usepackage{graphicx,color,dcolumn,booktabs,bm}
\usepackage{longtable,lscape}
\usepackage{txfonts}
\usepackage{overpic}
\usepackage{amssymb}
\usepackage{indentfirst}
\usepackage{feynmf}   %{feynmp}
\usepackage{slashed}  %for Feynman symbols
\usepackage{cases}
\usepackage{color}
\usepackage{multirow}
\usepackage{epstopdf}
\usepackage{graphicx,color,dcolumn,booktabs,bm}
\usepackage[colorlinks, citecolor=blue,anchorcolor=red,menucolor=red, linkcolor=red,filecolor=red,runcolor=red,urlcolor=blue,frenchlinks=red]{hyperref}

\graphicspath{{Figures/}} %

\begin{document}

\title{$\Xi_c(3055)$ as a scaling point to establish the excited $\Xi_c^{(\prime)}$ family}

\author{Xiao-Huang Hu$^1$}\thanks{These authors contribute equally to this work.}
\author{Zhe-Tao Miu$^2$}\thanks{These authors contribute equally to this work.}
\author{Zi-Xuan Ma$^2$}\thanks{These authors contribute equally to this work.}
\author{Qi Huang$^2$}\email{06289@njnu.edu.cn}
\author{Yue Tan$^3$}\email{tanyue@ycit.edu.cn }
\author{Jia-Lun Ping$^2$}\email{jlping@njnu.edu.cn}
\affiliation{
$^1$Department of Physics,Changzhou Institute of Industry Technology, Changzhou 213164, China\\
$^2$Department of Physics and Technology, Nanjing Normal University, Nanjing 210023, China\\
$^3$Department of Physics, Yancheng Institute of Technology, Yancheng 224000, China}

\begin{abstract}
%The low-lying orbital excited $\Xi_c^{(\prime)}$ baryons are studied in the chiral quark model combined with the $^3P_0$ decay mechanism. Motivated by the LHCb observation that $\Xi_c(3055)$ is a $\lambda$-mode $1D$ state with $J^P=3/2^+$, we take this state as a scaling point to constrain model parameters. Our calculations provide a consistent assignment for the observed $\Xi_c$ states below 3.1 GeV. The $\lambda$-mode $1P$ doublet $(\frac12^-,\frac32^-)$ corresponds to $\Xi_c(2790)$ and $\Xi_c(2815)$, while $\Xi_c(2923)$, $\Xi_c(2939)$, $\Xi_c(2965)$ and $\Xi_c(2882)$ are interpreted as $P$-wave excitations with mixed $\lambda/\rho$ modes. The $2S$ state $\Xi_c(\frac12^+)$ is identified with $\Xi_c(2970)$, and $\Xi_c(3055)$ together with $\Xi_c(3080)$ forms a $\lambda$-mode $D$-wave doublet $(\frac32^+,\frac52^+)$. Additionally, $\Xi_c(3123)$ may be a $\Sigma$-type $1D$ state. These results clarify the low-lying $\Xi_c$ spectrum and call for future experimental tests via $\Lambda D$ and $\Sigma D$ decay channels. 
The low-lying orbital excitations of $\Xi_c^{(\prime)}$ baryons are studied within the chiral quark model combined with the $^3P_0$ decay mechanism. Motivated by the LHCb observation that $\Xi_c(3055)$ is a $\lambda$-mode $1D$ state with $J^P=3/2^+$, we take it as a reference to constrain model parameters and perform a systematic analysis of states below 3.1 GeV.  Our results provide a consistent interpretation of the observed $\Xi_c$ spectrum: $\Xi_c(2790)$ and $\Xi_c(2815)$ are well described as a $\lambda$-mode $1P$ doublet, the structures around 2.9 GeV can be interpreted as $P$-wave excitations with mixed $\lambda/\rho$ modes, and $\Xi_c(2970)$ is consistent with a $2S$ assignment. Furthermore, $\Xi_c(3055)$ together with $\Xi_c(3080)$ forms a $\lambda$-mode $D$-wave doublet, while $\Xi_c(3123)$ favors a $1D$ assignment. These results provide a coherent picture of the low-lying $\Xi_c$ spectrum and call for further experimental tests.
\end{abstract}

\maketitle

\setcounter{totalnumber}{5}

\section{\label{sec:introduction}Introduction}

As one of the fundamental problems of strong interaction, decoding how interactions arise between quarks decides the follow-up behaviors of hadrons. Among the development history of the theory of strong interaction, the study of baryons, which is considered as a kind of hadron that composed by three constituent quarks, gave birth to color degree of freedom. Thus, studies on the baryon system must be a very interesting topic in hadron physics, not only because it will deepen our understanding on strong interaction, but also due to its property as the simplest few-body system, which can help us recognize further on the few-body interaction.

\begin{figure}[ht]
\begin{center}
\includegraphics[width=0.46\textwidth]{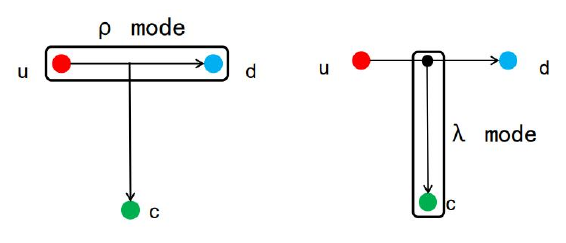} \vspace{-0.1in}
\caption{$\Xi_c^{(\prime)}$ system with $\lambda-$ or $\rho-$ mode excitations.}\label{mode}
\end{center}
\end{figure}

Among baryon systems, singly heavy baryons play a special role, as they offer an ideal platform to study the interplay between heavy quark symmetry and chiral symmetry. In particular, the $\Xi_c^{(\prime)}$ family, composed of one charm quark and two light quarks with different masses, provides a unique environment where both SU(3) flavor symmetry and chiral symmetry are explicitly broken, which provides a great platform to examine the inner Golstone boson exchanges. Within this system, excitations can be classified into two modes: the $\lambda$ mode, corresponding to the excitation between the heavy quark and the light diquark, and the $\rho$ mode, corresponding to the excitation within the light-quark subsystem, which are shown in Fig \ref{mode}.

Experimentally, a number of excited $\Xi_c^{(\prime)}$ states have been observed in recent years~\cite{ParticleDataGroup:2024cfk,BaBar:2007xtc,Belle:2018yob,BaBar:2007zjt,Belle:2016lhy,Belle:2008yxs,LHCb:2025mge}, as summarized in Table~\ref{1-4}. However, the quantum numbers of many of these states remain undetermined, especially for the structures around 2.9–3.1 GeV, where different theoretical approaches, including lattice QCD~\cite{Bahtiyar:2020uuj}, QCD sum rules~\cite{Chen:2016phw,Chen:2015kpa}, quark models~\cite{Ebert:2011kk,Roberts:2007ni,Yu:2022ymb,Li:2024zze}, and flux-tube models~\cite{Chen:2014nyo}, often lead to conflicting interpretations. This situation calls for further systematic studies.

Thus, a consistent interpretation of the observed spectrum is still lacking, and to study this project, a QCD-inspired quark model is still one of the most commonly used approaches. The chiral quark model, which incorporates Goldstone-boson exchanges between light quarks, has been successfully applied to describe the hardron spectra~\cite{Vijande:2004he,Segovia:2008zza,Segovia:2008zz,Ortega:2016hde}, hadron-hadron
interaction~\cite{Fernandez:1993hx,Valcarce:1994nr,Ortega:2016mms,Ortega:2016pgg} and multiquark structures~\cite{Vijande:2006jf,Yang:2020atz,Huang:2023jec}. Notably, in baryon sectors, the chiral quark model has been used to study the spectra and decays of excited $N$ and $\Sigma$ ~\cite{Tan:2025kjk,Yao:2025qor,Xiao:2015gra}, and successfully explaine the internal structures of several excited states, such as $N(1535)$, the Roper resonance $N(1440)$, and $\Sigma(1750)$. In addition, their results do indicate that Goldstone boson exchange plays a very important role in understanding the baryon properties. This makes it a suitable framework for the present study.

Recently, the LHCb Collaboration determined that $\Xi_c(3055)$ is a $\bar{3}_F$ $\lambda$-mode $1D$ state with $J^P=3/2^+$~\cite{LHCb:2024eyx}. Such result obviously provides an important reference point for constraining model parameters and organizing the $\Xi_c^{(\prime)}$ spectrum. Motivated by this, in this work we perform a systematic study of the low-lying $2S$, $1P$, and $1D$ states of $\Xi_c^{(\prime)}$ baryons within the chiral quark model combined with the $^{3}P_{0}$ decay mechanism. By comparing the calculated mass spectra and decay widths with experimental data, we aim to provide a coherent interpretation of the observed $\Xi_c^{(\prime)}$ states.

The paper is organized as follows. In Sec.~II, we briefly introduce the theoretical framework. Sec.~III presents the numerical results and discussions. Finally, a summary is given in Sec.~IV.

\begin{table*}[ht]
\caption{The basic experimental information of the excited $\Xi_{c}$ baryons  observed so far includes the $J^{P}$, mass, total width, and decay mode.\label{1-4}}
\setlength{\tabcolsep}{14pt}
\begin{tabular}{ccccccccc}
\hline \hline
State                   &$J^{P}$~\cite{ParticleDataGroup:2024cfk}   &Mass (MeV)~\cite{ParticleDataGroup:2024cfk}          &Width (MeV)~\cite{ParticleDataGroup:2024cfk}     &Decay modes~\cite{ParticleDataGroup:2024cfk}                     \\ \hline
$\Xi_{c}(2790)^{+}$ &$\frac{1}{2}^{-}$          &$2791.9\pm0.5$             &$8.9\pm1.0$               &$\Xi_{c}^{\prime}\pi,\Xi_{c}^{0}\gamma,\Xi_{c}^{+}\gamma$ \\
$\Xi_{c}(2790)^{0}$ &$\frac{1}{2}^{-}$          &$2793.9\pm0.5$             &$10.0\pm1.1$               &$\Xi_{c}^{\prime}\pi,\Xi_{c}^{0}\gamma,\Xi_{c}^{+}\gamma$                 \\
$\Xi_{c}(2815)^{+}$ &$\frac{3}{2}^{-}$          &$2816.51\pm0.25$             &$2.43\pm0.26$               &$\Xi_{c}^{\prime}\pi,\Xi_{c}(2645)\pi,\Xi_{c}^{0}\gamma,\Xi_{c}^{+}\gamma$ \\
$\Xi_{c}(2815)^{0}$ &$\frac{3}{2}^{-}$          &$2819.79\pm0.30$             &$2.54\pm0.25$               &$\Xi_{c}^{\prime}\pi,\Xi_{c}(2645)\pi,\Xi_{c}^{0}\gamma,\Xi_{c}^{+}\gamma$ \\
$\Xi_{c}(2923)^{0}$ &$?^{?}$~\cite{LHCb:2020iby}          &$2923.04\pm0.35$~\cite{LHCb:2020iby}             &$7.1\pm2.0$~\cite{LHCb:2020iby}               &$\Lambda_{c}K$~\cite{LHCb:2020iby} \\
$\Xi_{c}(2939)^{0}$ &$?^{?}$~\cite{LHCb:2020iby}          &$2938.55\pm0.30$~\cite{LHCb:2020iby}             &$10.2\pm1.4$~\cite{LHCb:2020iby}               &$\Lambda_{c}K$~\cite{LHCb:2020iby} \\
$\Xi_{c}(2965)^{0}$ &$?^{?}$~\cite{LHCb:2020iby}          &$2964.88\pm0.33$~\cite{LHCb:2020iby}             &$14.1\pm1.6$~\cite{LHCb:2020iby}               &$\Lambda_{c}K$~\cite{LHCb:2020iby} \\
$\Xi_{c}(2970)^{+}$ &$\frac{1}{2}^{+}$          &$2964.3\pm1.5$             &$20.9^{+2.4}_{-3.5}$               &$\Lambda_{c}K\pi,\Sigma_{c}(2455)K,\Xi_{c}^{\prime}\pi,\Xi_{c}(2645)\pi,\Xi_{c}\pi\pi$ \\
$\Xi_{c}(2970)^{0}$ &$\frac{1}{2}^{+}$          &$2967.1\pm1.7$             &$14.1\pm0.9\pm1.3$               &$\Lambda_{c}K\pi,\Sigma_{c}(2455)K,\Xi_{c}^{\prime}\pi,\Xi_{c}(2645)\pi,\Xi_{c}\pi\pi$ \\
$\Xi_{c}(3055)^{+}$ &$?^{?}$          &$3055.9\pm0.4$             &$7.8\pm1.9$               &$\Sigma_{c}(2455)K,\Lambda D$ \\
$\Xi_{c}(3080)^{+}$ &$?^{?}$          &$3077.2\pm0.4$             &$3.6\pm1.1$               &$\Lambda_{c}K\pi,\Sigma_{c}(2455)K,\Sigma_{c}(2520)K,\Lambda D$ \\
$\Xi_{c}(3080)^{0}$ &$?^{?}$          &$3079.9\pm1.4$             &$5.6\pm2.2$               &$\Lambda_{c}K\pi,\Sigma_{c}(2455)K,\Sigma_{c}(2520)K,\Lambda D$ \\
$\Xi_{c}(3123)^{+}$ &$?^{?}$          &$3122.9\pm1.3$             &$4\pm4$               &$\Sigma_{c}(2520)K$ \\
\hline \hline
\end{tabular}
\end{table*}

\begin{table*}[ht]
\renewcommand{\arraystretch}{1.5} % Adjust the 1.5 to your preferred amount
\setlength{\tabcolsep}{11pt}
\caption{Quantum number interpretations of excited $\Xi_{c}$ under different models.\label{46}}
\begin{tabular}{ccccccccc}
\hline \hline
Resonances     &HHCPT\cite{Cheng:2021qpd}    &RQM\cite{Ebert:2011kk}         &NQM\cite{Roberts:2007ni}   &QSR\cite{Chen:2016phw,Chen:2015kpa}     &RTF\cite{Chen:2014nyo}   &$^{3}P_{0}$\cite{Chen:2007xf}  &LQCD\cite{Bahtiyar:2020uuj}                \\ \hline
$\Xi_{c}(2790)$ &$\frac{1}{2}^{-}$          &$\frac{1}{2}^{-}$             &$\frac{1}{2}^{-}$               &$\frac{1}{2}^{-}$     &$\frac{1}{2}^{-}$        &$\frac{1}{2}^{-}$      &$\frac{1}{2}^{-}$                           \\
$\Xi_{c}(2815)$ &$\frac{3}{2}^{-}$          &$\frac{3}{2}^{-}$             &$\frac{3}{2}^{-}$               &$\frac{3}{2}^{-}$     &$\frac{3}{2}^{-}$        &$\frac{3}{2}^{-}$      &$\frac{3}{2}^{-}$                           \\
$\Xi_{c}(2923)$ &$-$          &$(\frac{1}{2},\frac{3}{2},\frac{5}{2})^{-}$             &$-$               &$\frac{1}{2}^{-}$     &$-$        &$-$          &$\frac{1}{2}^{-}$                 \\
$\Xi_{c}(2939)$ &$-$          &$(\frac{1}{2},\frac{3}{2},\frac{5}{2})^{-}$             &$-$               &$\frac{1}{2}^{-}$     &$-$        &$-$          &$\frac{1}{2}^{-}$                 \\
$\Xi_{c}(2965)$ &$-$          &$(\frac{1}{2},\frac{3}{2},\frac{5}{2})^{-}$             &$-$               &$\frac{1}{2}^{-}$     &$-$        &$-$          &$\frac{1}{2}^{-}$                 \\
$\Xi_{c}(2970)$ &$\frac{1}{2}^{+}$           &$\frac{1}{2}^{+}$             &$-$               &$\frac{1}{2}^{-}$   &$\frac{1}{2}^{+}$       &$D$-wave  &$(\frac{1}{2}^{\pm},\frac{3}{2}^{\pm},\frac{5}{2}^{\pm})$                         \\
$\Xi_{c}(3055)$ &$-$          &$\frac{3}{2}^{+}$             &$-$               &$\frac{3}{2}^{+}$     &$\frac{3}{2}^{+}$        &$-$     &$\frac{1}{2}^{-}$                                                     \\
$\Xi_{c}(3080)$ &$\frac{5}{2}^{+}$          &$\frac{5}{2}^{+}$             &$\frac{5}{2}^{+}$               &$\frac{5}{2}^{+}$     &$\frac{5}{2}^{+}$        &$D$-wave &$\frac{1}{2}^{+}$                           \\
$\Xi_{c}(3123)$ &$-$          &$\frac{7}{2}^{+}$             &$-$               &$-$     &$(\frac{1}{2},\frac{3}{2})^{-}$        &$-$       &$-$                    \\
\hline \hline
\end{tabular}
\end{table*}

\section{THEORETICAL FRAMEWORK}

\subsection{Chiral quark model}
As an application of the chiral symmetry, the chiral quark model contains kinematic term, confinement, one gluon exchange, and Goldstone boson exchange as

\begin{eqnarray}
H & =& \sum_{i=1}^{3}\left( m_i+\frac{p^2_i}{2m_i}\right)-T_{CM} +\sum_{j>i=1}^{3}\left[ V_{CON}(\boldsymbol{r}_{ij}) \right.\nonumber \\
&&\left.+V_{OGE}(\boldsymbol{r}_{ij})+V_{GBE}(\boldsymbol{r}_{ij}) \right] ,
\end{eqnarray}
where $m_i$ is the constituent mass of quark (antiquark), $\bf{p}_i$ is momentum of quark $i$, $T_{cm}$ is the kinetic energy of the center-of mass, $\boldsymbol{r}_{ij}$ means the relative coordinate between quark $i$ and $j$, and $V_{CON}$, $V_{OGE}$, $V_{GBE}$ are color confinement, one-gluon exchange potential, Goldstone boson exchange potentials. These three kinds of potentials reveal the most relevant features of QCD at low energy regime, i.e., color confinement, asymptotic freedom and chiral symmetry spontaneous breaking~\cite{Vijande:2004he}.

For color confinement potential, in this work, we adopt the linear form, which refers to the very famous Cornell potential as
\begin{eqnarray}
\setlength{\abovedisplayskip}{5pt}
\setlength{\belowdisplayskip}{5pt}
V^C_{CON}({{\bf r}_{ij}})  &=&  \boldsymbol{\lambda}_i^c\cdot \boldsymbol{\lambda}_j^c
   (-a_c r_{ij}-\Delta), \\
V^{SO}_{CON}({{\bf r}_{ij}})&=& -(\boldsymbol{\lambda}_i^c \cdot\boldsymbol{\lambda}_j^c )
 \frac{a_{c}r_{ij}}{4m_{i}^{2}m_{j}^{2}}
 [((m_{i}^{2}+m_{j}^{2})(1-2a_{s})\nonumber \\
&&+4m_{i}m_{j}(1-a_{s}))(\boldsymbol{S}_{+}\cdot\boldsymbol{L})\nonumber \\
&&+(m_{j}^{2}-m_{i}^{2})(1-2a_{s})(\boldsymbol{S}_{-}\cdot\boldsymbol{L})].
\end{eqnarray}
Here, $V^{SO}_{CON}({{\bf r}_{ij}})$ denotes the Thomas-precession effect, $a_c$ and $\Delta$ are model parameters, $\boldsymbol{S}_{\pm}=\boldsymbol{S}_{i}\pm\boldsymbol{S}_{j}$, and $\boldsymbol{\lambda}^c$ represent the SU(3) Gell-Mann matrices.

For one gluon exchange interaction, it contains the so-called coulomb and color-magnetism interactions, in addition with spin-orbit potential and tensor potenial, which arise from QCD perturbation effects and low-order relativistic corrections, as
\begin{eqnarray}
V^{C}_{OGE}({{\bf r}_{ij}})&=&  \frac{\alpha_s}{4} \boldsymbol{\lambda}_i^c \cdot \boldsymbol{\lambda}_j^c
   \left[ \frac{1}{r_{ij}}-\frac{\boldsymbol{\sigma}_i\cdot \boldsymbol{\sigma}_j}{6m_im_j}  \frac{e^{-r_{ij}/r_0(\mu)}}{r_{ij}r^2_0(\mu)}\right] ,      \\
V_{OGE}^{SO}({{\bf r}_{ij}})&=&  -\frac{\alpha_s}{16m_i^2m_j^2} \boldsymbol{\lambda}_i^c \cdot \boldsymbol{\lambda}_j^c \left[ \frac{1}{r_{ij}^3}- \frac{e^{-r_{ij}/r_g(\mu)}}{r_{ij}^3} \right.  \nonumber\\
&&\left.\times(1+\frac{r_{ij}}{r_g(\mu)} )\right]
    \times\left[((m_i+m_j)^2 +2m_im_j)\right.\nonumber\\
    &&\left.\times\left(\mathbf{S}_+ \cdot \mathbf{L}\right)  
    +(m_j^2-m_i^2)(\mathbf{S}_- \cdot \mathbf{L} )\right],  \\
V_{OGE}^T({{\bf r}_{ij}}) &=& -\frac{\alpha_s}{16 m_im_j} \boldsymbol{\lambda}_i^c \cdot \boldsymbol{\lambda}_j^c \left[ \frac{1}{r_{ij}^3}- \frac{e^{-r_{ij}/r_g(\mu)}}{r_{ij}}\right.\nonumber\\
&&\left.\times\left(\frac{1}{r_{ij}^2} +\frac{1}{3r_{g}^2(\mu)}+\frac{1}{r_{ij}r_g(\mu)}\right) \right]S_{ij}.
\end{eqnarray}
Here, $\mu$ is the reduced mass of two interacting quarks, $\boldsymbol{\sigma}$ represent the SU(2) Pauli matrices, $S_{ij}$ is the tensor operator, $r_0(\mu)\equiv\hat{r}_0/\mu$, $r_{g}(\mu)\equiv\hat{r}_g/\mu$, with $\hat{r}_0$ and $\hat{r}_g$ being parameters, and $\alpha_s$ denotes the effective flavor-dependent strong coupling constant.

Due to chiral symmetry breaking, Goldstone boson exchange potentials will appear between light quarks ($u$,$d$ and $s$), which have the expressions as
\begin{eqnarray}
V_{GBE}({{\bf r}_{ij}})&=&V_{\pi}({{\bf r}_{ij}})+V_{K}({{\bf r}_{ij}})+V_{\eta}({{\bf r}_{ij}})+V_{sc}({{\bf r}_{ij}}), \\
V_{\pi}({{\bf r}_{ij}})&=&\frac{g^2_{ch}}{4\pi}\frac{m^2_\pi}{12m_im_j}\frac{\Lambda^2_\pi m_\pi}{\Lambda^2_\pi-m^2_\pi}\sum_{a=1}^{3}\lambda_i^a \lambda_j^a \nonumber \\
&&\times\left\{(\boldsymbol{\sigma}_i \cdot \boldsymbol{\sigma}_j)\left[ Y(m_{\pi}r_{ij})- \frac{\Lambda^3_\pi}{m^3_\pi}Y (\Lambda_{\pi}r_{ij})\right]\right. \nonumber  \\
&&\left.+\left[H(m_{\pi}{\bf r}_{ij})-\frac{\Lambda^3_\pi}{m^3_\pi} H(\Lambda_{\pi}r_{ij})\right]S_{ij} \right\},    \\
V_{K}({{\bf r}_{ij}})&=&\frac{g^2_{ch}}{4\pi}\frac{m^2_K}{12m_im_j}\frac{\Lambda^2_K m_K}{\Lambda^2_K-m^2_K}\sum_{a=4}^{7}\lambda_i^a \lambda_j^a\nonumber \\
&&\times\left\{(\boldsymbol{\sigma}_i \cdot \boldsymbol{\sigma}_j)\left[ Y(m_{K}r_{ij}) -\frac{\Lambda^3_K}{m^3_K}Y (\Lambda_{K}r_{ij}) \right]  \right.  \nonumber\\
&&\left.+\left[H(m_{K}{\bf r}_{ij})-\frac{\Lambda^3_K}{m^3_K} H(\Lambda_{K}r_{ij})\right] S_{ij} \right\},    \\
V_{\eta}({{\bf r}_{ij}})&=&\frac{g^2_{ch}}{4\pi}\frac{m^2_\eta}{12m_im_j}\frac{\Lambda^2_\eta m_\eta}{\Lambda^2_\eta-m^2_\eta}\nonumber \\
&&\times\left[\cos\theta_{P}(\lambda_i^8 \lambda_j^8)-\sin\theta_{P}(\lambda_i^0 \lambda_j^0)\right]\nonumber \\
&&\times\left\{(\boldsymbol{\sigma}_i \cdot \boldsymbol{\sigma}_j)\left[ Y(m_{\eta}r_{ij})-\frac{\Lambda^3_\eta}{m^3_\eta}Y (\Lambda_{\eta}r_{ij}) \right]\right.\nonumber \\
&&\left.+\left[H(m_{\eta}{\bf r}_{ij})-\frac{\Lambda^3_\eta}{m^3_\eta} H(\Lambda_{\eta}r_{ij})\right] S_{ij} \right\}, \\
Y(x)&=&e^{-x}/x,\quad H(x)=\left(1+\frac{3}{x}+\frac{3}{x^2}\right)Y(x),
\end{eqnarray}
where $\lambda^a$ are the Gell-Mann matrices, $\Lambda_\chi~(\chi=\pi,K,\eta)$ are the cut-offs, $m_{\chi}$ are the masses of Goldstone bosons, and $g^2_{ch}$ is the chiral field coupling constant, which is determined from the $NN\pi$ interaction through
\begin{eqnarray}
\frac{g^2_{ch}}{4\pi}=\frac{9}{25}\frac{g^2_{\pi NN}}{4\pi}\frac{m^2_{u,d}}{m^2_N}.
\end{eqnarray}
Additionally, as the extension of the $\sigma$ meson exchange, the scalar nonet exchange $V_{sc}$ is also included as
\begin{eqnarray}
V_{sc}({{\bf r}_{ij}})&=&V_{a_0}({{\bf r}_{ij}})\sum_{a=1}^{3}
	\lambda_i^a \lambda_j^a+V_{\kappa}({{\bf r}_{ij}})\sum_{a=4}^{7}
	\lambda_i^a \lambda_j^a \nonumber \\
&&+ V_{f_0}({{\bf r}_{ij}})
	\lambda_i^8 \lambda_j^8+V_{\sigma}({{\bf r}_{ij}})\lambda_i^0 \lambda_j^0,   \\
V_{s}({{\bf r}_{ij}}) & =& -\frac{g^2_{ch}}{4\pi} \frac{\Lambda^2_s m_s}{\Lambda^2_s-m^2_s}
	\left[ Y(m_{s}r_{ij})-\frac{\Lambda_s}{m_s}Y(\Lambda_{s}r_{ij})\right] \nonumber  \\
&&+\frac{m^3_s}{2m_im_j}\left[G(m_{s}{r}_{ij})-\frac{\Lambda_{s}^{3}}{m_{s}^{3}}G(\Lambda_{s}{r}_{ij})\right]\nonumber\\
&&\times\mathbf{L}\cdot \mathbf{S}~~(s=a_0,\kappa,f_0,\sigma),\\
G(x)&=&\left(1+\frac{1}{x}\right)\frac{Y(x)}{x},
\end{eqnarray}
with $m_s$ and $\Lambda_s$ are the mass and cutoff that correspond to the scalar meson $s$, respectively.

To determine the model parameters of our work, we perform a combined fit on the well established ground mesons and baryons that relative to our next calculations. The model parameters are listed in Table~\ref{4-0}, and the fitted baryon and meson masses are presented in the Table~\ref{4-3} along with
the experimental values. Since it is difficult to use the same set of parameters to simultaneously describe the baryon and meson spectra well, we use different values of $\alpha_s$ for quark-quark and quark-antiquark pairs. As we can see from Table~\ref{4-3}, most of our fitted masses are close to the experimental values. We want to emphasize here that for the purpose of unification, the flavor functions of all the states appeared in this work are described in the SU(3) symmetry. However, during the fit, we surprisingly find some possible competitions between one-gluon-exchange and one-boson-exchange potentials, which will break the common sense on quantum chromodynamics that $\alpha_s$ will decrease with the momentum transfer, as Table~\ref{4-0} shows. For the discussion on this issue, we will present it in detail in the appendix.
\begin{table}[!htb]
\centering
\caption{Quark model parameters used in the calculation, where the numbers with superscript "$\ast$" means they are previously fixed during the fit.\label{4-0}}
\begin{tabular}{c|cc c}
 \hline \hline
                   &$m_{u,d}$ (MeV)   &~~~~313$^\ast$ & \\
  Quark masses      &$m_s$ (MeV)  &~~~~555$^\ast$ & \\
                    &$m_c$ (MeV)  &~~~~1800$^\ast$ & \\  \hline
                   &$\Lambda_\pi$ (fm$^{-1}$)  &~~~~4.20$^\ast$ & \\
                   &$\Lambda_{\eta,K}$ (fm$^{-1}$)      &~~~~5.20$^\ast$ & \\
                   &$m_\pi$ (fm$^{-1}$)  &~~~~0.70$^\ast$ & \\
Goldstone bosons   &$m_K$ (fm$^{-1}$)  &~~~~2.51$^\ast$ & \\
                   &$m_\eta$ (fm$^{-1}$)  &~~~~~2.77$^\ast$ & \\
                   &$g^2_{ch}/(4\pi)$  &~~~~0.54$^\ast$ & \\
                   &$\theta_P(^\circ)$  &~~~~-15$^\ast$ & \\  \hline
                   &$a_c$ (MeV$\cdot$fm$^{-1}$)  &~~~~108.34$\pm$0.16 & \\
     Confinement
                    &$\Delta$ (MeV)  &~~~-82.99$\pm$0.15 & \\
                    &$a_{s}$     &~~~~0.777$^\ast$ & \\ \hline
                   &$m_\sigma$ (fm$^{-1}$)  &~~~~3.42$^\ast$ & \\
                   &$\Lambda_\sigma$ (fm$^{-1}$)  &~~~~4.20$^\ast$ & \\
Scalar nonet       &$\Lambda_{a_0,\kappa,f_0}$ (fm$^{-1}$)  &~~~~5.20$^\ast$ & \\
                   &$m_{a_0,\kappa,f_0}$ (fm$^{-1}$)  &~~~~4.97$^\ast$ & \\  \hline
                 &$\hat{r}_0=\hat{r}_g~$(MeV$\cdot$fm)  &~~~~24.7$^\ast$ & \\
                    &$\alpha_{uu}$  &~~~~0.477$\pm$0.0014 & \\
               &$\alpha_{us}$  &~~~~0.566$\pm$0.0017 & \\
                    &$\alpha_{ss}$  &~~~~0.502$\pm$0.0015 & \\
      OGE                &$\alpha_{uc}$ &~~~~0.713$\pm$0.0011 & \\
                    &$\alpha_{sc}$  &~~~~0.668$\pm$0.0010 & \\ 
                     &$\alpha_{u\bar{u}}$  &~~~~0.566$\pm$0.0015 & \\
                      &$\alpha_{u\bar{s}}$  &~~~~0.534$\pm$0.0013 & \\
               &$\alpha_{s\bar{s}}$  &~~~~0.395$\pm$0.0015 & \\&$\alpha_{u\bar{c}}$  &~~~~0.626$\pm$0.0009 & \\
                    &$\alpha_{s\bar{c}}$  &~~~~0.530$\pm$0.0008 & \\  \hline \hline
\end{tabular}
\end{table}

\begin{table}[h]
\centering
\caption{The masses of ground state baryons and mesons in unit of MeV.\label{4-3}}
\begin{tabular}{ccccccccccccccccccc}
\hline \hline
~     &N~       &$\Delta$~  &$\Lambda$~     &$\Sigma$~          &$\Sigma^{*}$~                                                     \\ \hline
CQM~ &940$\pm$10     &1285$\pm$7   &1119$\pm$8          &1143$\pm$9            &1385$\pm$7                                                                                                                                     \\
Exp.~ &939    &1232      &1116~          &1189~             &1383~                                                                                    \\ \hline

 ~   &$\Xi$~  &$\Xi^{*}$  &$\Omega$~  &$\Lambda_c$~      &$\Sigma_{c}$~        \\ \hline
CQM~    &1350$\pm$7    &1501$\pm$6  &1656$\pm$6~  &2274$\pm$7~    &2475$\pm$6~                          \\
Exp.~   &1318~ &1533  &1672~  &2286~     &2455~                                             \\ \hline

~ &$\Sigma_{c}^{*}$~    &$\Xi_c$~  &$\Xi_c^{\prime}$~ &$\Xi_{c}^{\prime*}$~ &$\Omega_{c}$~                   \\ \hline
CQM~ &2541$\pm$6~         &2473$\pm$7~  &2565$\pm$6~       &2637$\pm$6~    &2689$\pm$5~                  \\
Exp.~   &2520~         &2471~ &2579~    &2645~    &2695~       
\\ \hline
~    &$\Omega_{c}^{*}$   &$\pi$~   &$\rho$~   &$K$~  &$K^{*}$                                 \\ \hline
CQM~  &2765$\pm$5~    &140$\pm$8~     &821$\pm$3~&494$\pm$4~   &975$\pm$2~        \\
Exp.~ &2770~  &140~     &775~        &495~    &892~       \\ \hline

~       &$\omega$~  &$\eta^{\prime}$~ &$\phi$~ &$D$~    &$D^{*}$~          \\ \hline
CQM~      &701$\pm$3~ &958$\pm$5~  &1040$\pm$5~ &1863$\pm$7~     &2063$\pm$4~   \\      
Exp.~  &782~ &958~  &1020~ &1864~     &2007~        
\\ \hline
~         &$D_s$~    &$D^{*}_s$~            \\ \hline
CQM~      &1967$\pm$5~     &2203$\pm$2~ \\      
Exp.~  &1968~     &2112~         
\\ \hline
\hline \hline
\end{tabular}
\end{table}

\subsection{Method to calculate mass spectra}

To correspond to the intrinsic chiral symmetry within our model, the flavor wave functions of $\Xi_c^{(\prime)}$ is constructed under the SU(3) symmetry, although the mass of $u/d$- and $s$-quarks are different. In addition, due to the much larger mass of the charm quark, chiral symmetry is approximately obeyed at this time. Thus, $\Xi_c$ can belong to the $\bar{3}_F$ represention as
\begin{eqnarray}
    |\Xi_c\rangle_f = \frac{1}{\sqrt{2}}(us-su)c,
\end{eqnarray}
while $\Xi_c^\prime$ can be accounted into the $6_F$ representation as
\begin{eqnarray}
    |\Xi_c^\prime\rangle_f = \frac{1}{\sqrt{2}}(us+su)c,
\end{eqnarray}

Then, taken into account the anti-symmetric color wave function of any baryon as
\begin{eqnarray}
    |\Xi_c^{(\prime)}\rangle_c = \frac{1}{\sqrt{6}}(rgb + gbr + brg - grb - bgr - rbg),
\end{eqnarray}
under $S-L$ coupling representation, the total wave function of the $\Xi_c^{(\prime)}$ with total angular momentum $J$ can be represented as
\begin{eqnarray}
    |\Xi_c^{(\prime)}\rangle_J = |\Xi_c^{(\prime)}\rangle_c \otimes |\Xi_c^{(\prime)}\rangle_f \otimes \left[|\Xi_c^{(\prime)}\rangle_S \otimes |\Xi_c^{(\prime)}\rangle_L\right]_J,
\end{eqnarray}
where $\Xi_c^{(\prime)}\rangle_S$ and $|\Xi_c^{(\prime)}\rangle_L$ are spin and spatial wave functions that expanded as
\begin{eqnarray}
    &&|\Xi_c^{(\prime)}\rangle_S = \left[\left[\psi_{S_1}^{\rm spin}\otimes\psi_{S_2}^{\rm spin}\right]_{S_\rho}\otimes \psi_{S_3}^{\rm spin}\right]_{S},\\
    &&|\Xi_c^{(\prime)}\rangle_L = \left[\psi_{L_{\rho}}^{\rm orbit}\otimes\psi_{L_{\lambda}}^{\rm orbit}\right]_{L}.
\end{eqnarray}
Here, $S_\rho$ means the total spin of the two light quarks in the $\Xi_c^{(\prime)}$ baryon, $S$ is the total spin, $L_{\rho}$ is the relative orbital angular momentum between the two light quarks, $L_\lambda$ is the relative orbital angular momentum between the charm quark and the light quark pair, and $L$ is the total angular momentum.

For the spatial wave functions, we adopt the Rayleigh-Ritz variational method to solve the eigenvalue problem, where a basis expansion of the trial spatial wave function is performed. In this work, we select the widely used Gaussian expansion method (GEM)~\cite{Hiyama:2003cu} to expand each relative motion in the system. The GEM has proven to be an accurate and universal few-body calculation method.~\cite{Hu:2020zwc,Yang:2020fou,Tan:2020cpu}, whose critical point is to expand the radial part of the orbital wave function with a set of Gaussian functions as
\begin{eqnarray}
\psi^{\rm orbit}_l = \sum\limits_{n=1}^{n_{max}} C_n N_{nl}^r e^{-\nu_n r^2}\mathcal{Y}_l(\boldsymbol{r})
\end{eqnarray}
where $\mathcal{Y}_l(\boldsymbol{r})$ is the solid spherical harmonic function, $N_{nl}$ is the normalization constant,
\begin{eqnarray}
\emph{N}_{nl}=\left(\frac{2^{l+2}(2\nu_{n})^{l+3/2}}{\sqrt\pi(2l+1)!!}\right)^{\frac{1}{2}},
\end{eqnarray}
and $C_{n}$ are the variational parameters, which are determined by the dynamics of system. The Gaussian size are chosen as the following geometric progression
\begin{eqnarray}
\nu_{n}=\frac{1}{r^{2}_{n}}, \quad r_{n}=r_{min}\left(\frac{r_{max}}{r_{min}}\right)^{\frac{n-1}{n_{max}-1}},
\end{eqnarray}
where $r_{min}$ and $r_{max}$ are parameters, $n_{max}$ is the number of Gaussian functions, which are determined by the convergence of the results. In this work, we find that after setting $r_{min}=0.1$ fm, $r_{max}=2$ fm, and $n_{max}=8$, the spectra of the low-lying excited $\Xi_c^{(\prime)}$ baryons can be enoughly convergent.

Finally, to use the language of heavy quark symmetry, a representation transformation is performed, which is explicitly expressed as
\begin{eqnarray}
    \left[\psi_{J_l}\otimes\psi_{S_Q}^{\rm spin}\right]_J &=&(-1)^{L+S_\rho+J+\frac{1}{2}} \sum\limits_S\sqrt{2 J_l+1}\nonumber\\
    &&\times\sqrt{2 S+1}\left\{\begin{array}{ccc}
    L & S_\rho & J_l \\
    S_Q & J & S
    \end{array}\right\}\nonumber\\
    &&\times\left[\psi_S^{\rm spin} \otimes \psi_L^{\rm orbit}\right]_J.
\end{eqnarray}
Here, $S_Q$ is the spin of the heavy quark, $J_l = L \otimes S_\rho$ is the light degree of freedom, which is a good quantum number in heavy quark symmetry. Actually, under heavy quark limit, the heavy quark spin $S_Q$ is approximately freezed, which makes the interactions between the heavy quark and the two light quarks are strongly suppressed by heavy quark mass. This indicates that, for a singly heavy baryon, the state constructed by the $J_l-S_Q$ coupling may be closer to the physical state than that are composed by the $S-L$ scheme, although the masses may be different under different schemes. 

\subsection{Two body decay mechanism}

After calculating the mass spectra, we use the $^{3}P_{0}$ model to calculate the two body decays of the initial $\Xi_c^{(\prime)}$ baryons since under quark level, it has been widely applied and proved strongly effective to the studies of strong decay problems of hadrons \cite{Lu:2018utx,Chen:2007xf,Liu:2009fe} that allowed by the Okubo-Zweig-Iizuka rule. This model was first proposed by Micu in 1969 \cite{Micu:1968mk}, and was further developed by Le Yaounanc, Ackleh and Roberts \cite{LeYaouanc:1988fx,Ackleh:1996yt,Roberts:1992esl}, whose corresponding applicable scope has also been expanded from traditional hadrons to multi-quark states.

\begin{figure}[ht]
\begin{center}
\includegraphics[width=0.46\textwidth]{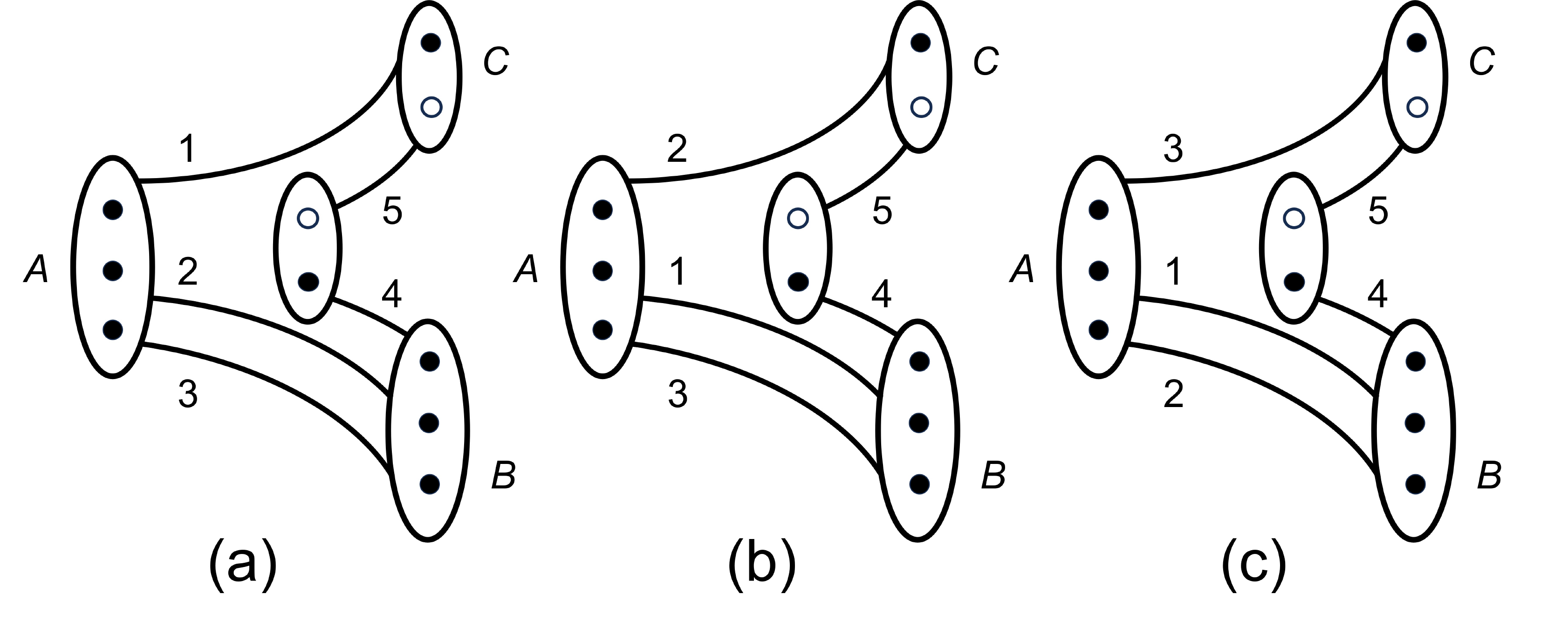} \vspace{-0.1in}
\caption{Baryon decay process of $A\rightarrow B+C$ in the $^{3}P_{0}$ model, taken from our previous work~\cite{Xie:2025gom}.\label{3p0}}
\end{center}
\end{figure}

According to this model, a pair of quarks with $J^{PC}=0^{++}$ is created firstly from the vacuum. Then, by the quark rearrangement process, this new $q\bar{q}$ pair from the vacuum combines with the three quarks within the initial baryon, resulting in the creation of the outgoing meson and baryon. When considering a baryon $A$ decays to a baryon $B$ and meson $C$,
there are usually three types of recombination, as shown in Fig. \ref{3p0}~\cite{Xie:2025gom}.

The original transition operator of the decay process $A\rightarrow B+C$ through $^{3}P_{0}$ model can be expressed under the momentum representation as~\cite{Chen:2024ukv},

\begin{eqnarray} \label{3-13}
  T&=&-3\gamma\sum_m\langle 1m;1-m|00\rangle\int
  d\mathbf{p}_4d\mathbf{p}_5\delta(\mathbf{p}_4+\mathbf{p}_5)\nonumber\\
  &&\times{\cal{Y}}^m_1\left(\frac{\mathbf{p}_4-\mathbf{p}_5}{2}\right)
  \chi_{45}^{1-m}\phi_{45}^0\omega_{45}^0b^\dagger_4(\mathbf{p}_4)d^\dagger_5(\mathbf{p}_5),
 \end{eqnarray}
where $\gamma$ describes the probability of creating a quark-antiquark pair with momenta $\mathbf{p}_4$ and $\mathbf{p}_5$ from the vacuum,respectively. The solid harmonic polynomial ${\cal{Y}}^m_1=|p|Y_l^m(\theta_{p},\phi_{p})$ reflects the $P$-wave distribution of the created quark pair in the momentum space. $\phi^{0}_{45}=(u\bar{u}+d\bar{d}+s\bar{s})/ \sqrt{3}$, $\omega^{0}_{45}=(r\bar{r}+g\bar{g}+b\bar{b})/ \sqrt{3}$, and $\chi^{1-m}_{45}$ are the flavor, color, and spin wave functions of the created quark pair, respectively.

However, when this original transition operator was used to calculate the mass shift of light mesons, a very large mass shift that beyond expectation would occur~\cite{Chen:2017mug}.
To avoid this problem, in Ref.~\cite{Chen:2017mug,Huang:2023jec}, convergence factors $e^{-\frac{r^2}{4f^2}}$ and damping factor $e^{-\frac{R^2_{AV}}{R^2_0}}$ are introduced, which reflect the cutoffs on the momentum and distance, i.e., the created quark pair should not have too large momentum, and the position it creates should not be too far away from the initial hadron. Thus, after doing a Fourier transformation on the original transition operator and multiplying the above two form factors, the transition operator can be modified as~\cite{Chen:2017mug}
\begin{eqnarray}
\label{3-14}
&&T= -3\gamma\sum_{m}\langle 1m;1-m|00\rangle\int
d\mathbf{r_4}d\mathbf{r_5}\left(\frac{1}{2\pi}\right)^{\frac{3}{2}}\nonumber \\
  &&\quad~~~\times 2^{-\frac{5}{2}}f^{-5} \mathcal{Y}_{1}^{m}(\mathbf{r})
  {\rm e}^{-\frac{\mathbf{r}^2}{4f^2}}
 {\rm e}^{-\frac{R_{AV}^2}{R_0^2}}\chi^{1-m}_{45}\nonumber \\
 &&\quad~~~\times\phi^{0}_{45}
 \omega^{0}_{45}b_4^{\dagger}(\mathbf{r_4})d_5^{\dagger}(\mathbf{r_5}),
\end{eqnarray}
with $\mathbf{r_4}$, $\mathbf{r_5}$, and $\mathbf{r}$ are the coordinates of the vacuum created quark, anti-quark, and the relative coordinate between them, respectively. 

Then, the matrix element for the transition $A\rightarrow B+C$ can be calculated as
\begin{eqnarray}
\label{3-8}
\langle
BC|T|A\rangle=\delta^3\left(\mathbf{P}_A-\mathbf{P}_B-\mathbf{P}_C\right){\cal{M}}^{M_{J_A}M_{J_B}M_{J_C}},
\end{eqnarray}
where~$\mathbf{P}_B$,~$\mathbf{P}_C$~are the momenta of the final $B$ and $C$ hadrons in the center-of-mass frame of baryon $A$. $\mathcal{M}^{M_{J_A}M_{J_B}M_{J_C}}$ is the helicity amplitude as
\begin{eqnarray}
&&\mathcal{M}^{M_{J_A} M_{J_B} M_{J_C}}(A \rightarrow B C)\nonumber\\
&&\quad= \sqrt{8 E_A E_B E_C} \sum_{\substack{M_{L_A}, M_{S_A},\\M_{L_B}, M_{S_B},\\M_{L_C}, M_{S_C},m}} \langle 1 m 1-m \mid 00\rangle\nonumber\\
&&\quad~~~\times\left\langle\chi_{S_C M_{S_C}}^{235} \chi_{S_B M_{S_B}}^{14} \mid \chi_{S_A M_{S_A}}^{123} \chi_{1-m}^{45}\right\rangle \nonumber\\
&&\quad~~~\times\left\langle L_A M_{L_A} S_A M_{S_A} \mid J_A M_{J_A}\right\rangle\nonumber \\
&&\quad~~~\times\left\langle L_B M_{L_B} S_B M_{S_B} \mid J_B M_{J_B}\right\rangle\nonumber\\
&&\quad~~~\times\left\langle L_C M_{L_C} S_C M_{S_C} \mid J_C M_{J_C}\right\rangle\nonumber\\
&&\quad~~~ \times\left\langle\varphi_C^{235} \varphi_B^{14} \mid \varphi_A^{123} \varphi_0^{45}\right\rangle I_{M_{L_B}, M_{L_C}}^{M_{L_A}, m},
\end{eqnarray}
\iffalse
\begin{small}
    \begin{align}
    &\mathcal{M}^{M_{J_A} M_{J_B} M_{J_C}}(A \rightarrow B C)\notag\\
    = &\sqrt{8 E_A E_B E_C}  \sum_{}\left\langle L_{\rho_A} M_{L_{\rho_A}} L_{\lambda_A} M_{L_{\lambda_A}} \mid L_A M_{L_A}\right\rangle\notag \\
    &\times\left\langle S_1 M_{S_1} S_2 M_{S_2} \mid S_{\rho_A} M_{S_{\rho_A}}\right\rangle
    \left\langle S_{\rho_A} M_{S_{\rho_A}} L_A M_{L_A} \mid J_{l_A} M_{J_{l_A}}\right\rangle\notag\\
    &\times\left\langle J_{l_A} M_{J_{l_A}} S_3 M_{S_3}\mid J_A M_{J_A}\right\rangle\left\langle L_{\rho_C} M_{L_{\rho_C}} L_{\lambda_C} M_{L_{\lambda_C}} \mid L_C M_{L_C}\right\rangle\notag\\
    &\times\left\langle S_2 M_{S_2} S_5 M_{S_5} \mid S_{\rho_C} M_{S_{\rho_C}}\right\rangle
    \left\langle S_{\rho_C} M_{S_{\rho_C}} L_C M_{L_C} \mid J_{l_C} M_{J_{l_C}}\right\rangle\notag\\
    &\times\left\langle J_{l_C} M_{J_{l_C}} S_3 M_{S_3}\mid J_C M_{J_C}\right\rangle\left\langle S_{1} M_{S_1} S_4 M_{S_4}\mid S_B M_{S_B}\right\rangle\notag\\
    &\times\left\langle L_{B} M_{L_{B}} S_B M_{S_B}\mid J_B M_{J_B}\right\rangle \langle S_4 M_{S_4} S_5 M_{S_5}\mid 1 -m\rangle\notag\\
    &\times \langle 1 m 1-m \mid 00\rangle \left\langle\varphi_C^{235} \varphi_B^{14} \mid \varphi_A^{123} \varphi_0^{45}\right\rangle I_{M_{L_B}, M_{L_C}}^{M_{L_A}, m},
    \end{align}
\end{small}
\fi
with the spatial integral $I_{M_{L_B}, M_{L_C}}^{M_{L_A}, m}$ can be written as
\begin{eqnarray}
&&I_{M_{L_B}, M_{L_C}}^{M_{L_A}, m}\notag\\
&&\quad= -3i\gamma\left(\frac{1}{2\pi}\right)^3 2^{-\frac{5}{2}}f^{-5}\int \psi^*_{L_{C}M_{L_{C}}}\left(\boldsymbol{r}_C,\boldsymbol{R}_C\right)\notag\\
&&\quad~~~\times\psi^*_{L_{B}M_{L_{B}}}(\boldsymbol{r}_B) \mathcal{Y}_{1}^{m}(\boldsymbol{r}_\nu)e^{-i\boldsymbol{r}_{BC}\cdot \boldsymbol{p}} e^{-\frac{r^2_v}{4f^2}}e^{\frac{-R^2_{AV}}{R^2_0}}\notag\\
&&\quad~~~\times \psi_{L_{A}M_{L_{A}}}(\boldsymbol{r}_A,\boldsymbol{R}_A) \mathrm{d} \boldsymbol{r}_A \mathrm{d} \boldsymbol{R}_A \mathrm{d} \boldsymbol{r}_\nu \mathrm{d} \boldsymbol{R}_{AV}.
\end{eqnarray}
Here, $E_i$ is the on-shell energy of hadron $i$, $\psi_{LM_{L}}$ is the spatial wave function of hadrons in coordinate representation, $\left\langle\chi_{S_C M_{S_C}}^{235} \chi_{S_B M_{S_B}}^{14} \mid \chi_{S_A M_{S_A}}^{123} \chi_{1-m}^{45}\right\rangle$ and $\left\langle\varphi_C^{235} \varphi_B^{14} \mid \varphi_A^{123} \varphi_0^{45}\right\rangle$ denote the spin and flavor overlap matrix element, respectively, $\boldsymbol{r}_A$ and $\boldsymbol{r}_C$ are the relative coordinates between two quarks in baryons $A$ and $C$, $\boldsymbol{R}_A$ and $\boldsymbol{R}_C$ are the relative coordinates between the third quark and the remaining quark pair in baryons $A$ and $C$, $\boldsymbol{r}_B$ represents the relative coordinate between the quark and anti-quark in the meson, $\boldsymbol{r}_\nu$ is the relative coordinate between the created quark pair from the vacuum, $\boldsymbol{R}_{AV}$ is the relative coordinate between the baryon $A$ and the vacuum created quark pair, $\boldsymbol{r}_{BC}$ is the the relative coordinate between the final $B$ and $C$, and $\boldsymbol{p}$ represents the momentum of the final hadron $B$ or $C$ in the center-of-mass frame of $A$, whose modulus is
\begin{align}
\left|\boldsymbol{p}\right|=\frac{\sqrt{[M_{A}^{2}-(M_{B}+M_{C})^{2}][M_{A}^{2}-(M_{B}-M_{C})^{2}]}}{2M_{A}},
\end{align}
with $M_{A}$, $M_{B}$, $M_{C}$ being the masses of the hadrons $A$, $B$, $C$, respectively.

Finally, the hadronic decay width of a process $A \rightarrow B + C$ is calculated as follows
\begin{align}
\Gamma =\pi^{2}\frac{\left|\boldsymbol{p}\right|}{M_{A}^{2}}\frac{\mathcal{S}}{2J_A +1}\sum_{M_{J_A},M_{J_B},M_{J_C}}\left|\mathcal{M}^{M_{J_A} M_{J_B} M_{J_C}}\right|^{2},
\end{align}
where $\mathcal{S}$ is a statistical factor for identical particles as $\mathcal{S}  \equiv 1/(1+\delta_{BC})$.

As for the $\gamma$, $f$, and $R_0$, in this work, we take the same value as Ref.~\cite{Chen:2017mug}, which can both reproduce the decay width of $\rho \to \pi \pi$ process and avoid the mass shift problem, which are
\begin{align}
  \gamma=32,~~~~~~ f=0.2~\mathrm{fm},~~~~~~ R_0=1.0~\mathrm{fm}.
\end{align}

\section{Results and discussions}
In this section, we present the mass spectra of the 2$S$, 1$P$ and 1$D$ $\Xi_c^{(\prime)}$ states calculated through the chiral quark model, as well as the corresponding decay widths under the $^{3}P_{0}$ decay mechanism.

The experimental value of several excited baryons with high status are listed in Table~\ref{1-4} and the mass spectra of $\Xi_c$ and $\Xi_c^{\prime}$ with the corresponding quantum numbers are shown in Table~\ref{4-1} and Table~\ref{4-2}, respectively, where $n_\lambda$, $L_\rho$ and $S_\rho$ denote the radial quantum number, relative orbital angular momentum and the total spin of the two light quarks ($us$), respectively. $L$ is the total orbital angular momentum of the baryon, $J_l$ is the light degree of freedom, and $J$ denotes the total angular momentum.

\begin{table*}[ht]
\renewcommand{\arraystretch}{1.5}
 \caption{The quantum number of $\Xi_{c}$ baryons in the $1S$, $2S$, $1P$ and $1D$ states and the corresponding masses in unit of MeV.\label{4-1}}
 \begin{tabular}{cccccccc|cccccccccccc}
 \hline
 \hline
 State~~~~&  Mass~~~~&$L_{\rho}$~~~~ & $L_{\lambda}$~~~~ & $L$ ~~~~& $S_{\rho}$~~~~ &$J_{l}$~~~~&$J$~~~~&State~~~~& Mass~~~& $L_{\rho}$ ~~~~& $L_{\lambda}$ ~~~~& $L$ ~~~~& $S_{\rho}$ ~~~~&$J_{l}$~~~~&$J$ \\
 \hline
 $\Xi_{c}(\frac{1}{2}^{+},1S)$&  2473$\pm$7&0&0&0&0&0&$\frac{1}{2}$& $\Xi_{c2}(\frac{3}{2}^{+},1D)$& 3043$\pm$8  &0&2&2&0&2&$\frac{3}{2}$ \\
 $\Xi_{c}(\frac{1}{2}^{+},2S)$&   2970$\pm$8  &0&0&0&0&0&$\frac{1}{2}$&  $\Xi_{c2}(\frac{5}{2}^{+},1D)$ &3078$\pm$8 &0&2&2&0&2&$\frac{5}{2}$\\
 %\hline
 $\Xi_{c1}(\frac{1}{2}^{-},1P)$&2783$\pm$7  &0&1&1&0&1&$\frac{1}{2}$&  $\Xi_{c2}(\frac{3}{2}^{+},1D)$&3353$\pm$9  &2&0&2&0&2&$\frac{3}{2}$\\
 $\Xi_{c1}(\frac{3}{2}^{-},1P)$&2830$\pm$7  &0&1&1&0&1&$\frac{3}{2}$ &$\Xi_{c2}(\frac{5}{2}^{+},1D)$&3373$\pm$9  &2&0&2&0&2&$\frac{5}{2}$ \\
 $\Xi_{c0}(\frac{1}{2}^{-},1P)$&2950$\pm$4  &1&0&1&1&0&$\frac{1}{2}$ &$\Xi_{c1}(\frac{1}{2}^{+},1D)$&3217$\pm$8  &1&1&2&1&1&$\frac{1}{2}$ \\
 $\Xi_{c1}(\frac{1}{2}^{-},1P)$&2910$\pm$3  &1&0&1&1&1&$\frac{1}{2}$& $\Xi_{c1}(\frac{3}{2}^{+},1D)$& 3255$\pm$4 &1&1&2&1&1&$\frac{3}{2}$\\
 $\Xi_{c1}(\frac{3}{2}^{-},1P)$&3057$\pm$5  &1&0&1&1&1&$\frac{3}{2}$&  $\Xi_{c2}(\frac{3}{2}^{+},1D)$& 3237$\pm$4 &1&1&2&1&2&$\frac{3}{2}$\\
 $\Xi_{c2}(\frac{3}{2}^{-},1P)$&3037$\pm$5  &1&0&1&1&2&$\frac{3}{2}$&  $\Xi_{c2}(\frac{5}{2}^{+},1D)$& 3303$\pm$6 &1&1&2&1&2&$\frac{5}{2}$\\
$\Xi_{c2}(\frac{5}{2}^{-},1P)$&3116$\pm$8  &1&0&1&1&2&$\frac{5}{2}$&  $\Xi_{c3}(\frac{5}{2}^{+},1D)$&3289$\pm$6  &1&1&2&1&3&$\frac{5}{2}$\\
 &&&&&&&&  $\Xi_{c3}(\frac{7}{2}^{+},1D)$& 3334$\pm$9 &1&1&2&1&3&$\frac{7}{2}$\\
\hline
 \hline
 \end{tabular}
 \end{table*}

\begin{table*}[ht]
\renewcommand{\arraystretch}{1.5}
 \caption{The quantum number of $\Xi_{c}^{\prime}$ baryons in the $1S$, $2S$, $1P$ and $1D$ states and the corresponding masses in unit of MeV.\label{4-2}}
 \begin{tabular}{cccccccc|cccccccccccc}
 \hline
 \hline
 State~~~~&  Mass~~~~& $L_{\rho}$~~~~ & $L_{\lambda}$~~~~ & $L$ ~~~~& $S_{\rho}$~~~~ &$J_{l}$~~~~&$J$~~~~&State~~~~& Mass~~~& $L_{\rho}$ ~~~~& $L_{\lambda}$ ~~~~& $L$ ~~~~& $S_{\rho}$ ~~~~&$J_{l}$~~~~&$J$ \\
 \hline
 $\Xi_{c}^{\prime}(\frac{1}{2}^{+},1S)$&2565$\pm$6    &0&0&0&1&1&$\frac{1}{2}$&$\Xi_{c1}(\frac{1}{2}^{+},1D)$&   3136$\pm$8   &0&2&2&1&1&$\frac{1}{2}$  \\
 $\Xi_{c}^{*}(\frac{3}{2}^{+},1S)$&2637$\pm$6&0&0&0&1&1&$\frac{3}{2}$&  $\Xi_{c1}(\frac{3}{2}^{+},1D)$& 3143$\pm$4      &0&2&2&1&1&$\frac{3}{2}$\\
 $\Xi_{c}^{\prime}(\frac{1}{2}^{+},2S)$& 3054$\pm$8 &0&0&0&1&1&$\frac{1}{2}$&  $\Xi_{c2}(\frac{3}{2}^{+},1D)$& 3172$\pm$4   &0&2&2&1&2&$\frac{3}{2}$ \\
 $\Xi_{c}^{*}(\frac{3}{2}^{+},2S)$& 3101$\pm$8&0&0&0&1&1&$\frac{3}{2}$& $\Xi_{c2}(\frac{5}{2}^{+},1D)$& 3153$\pm$5    &0&2&2&1&2&$\frac{5}{2}$\\
 %\hline
 $\Xi_{c0}^{\prime}(\frac{1}{2}^{-},1P)$&2874$\pm$4  &0&1&1&1&0&$\frac{1}{2}$&  $\Xi_{c3}(\frac{5}{2}^{+},1D)$& 3170$\pm$5   &0&2&2&1&3&$\frac{5}{2}$\\
 $\Xi_{c1}^{\prime}(\frac{1}{2}^{-},1P)$&2908$\pm$4  &0&1&1&1&1&$\frac{1}{2}$ & $\Xi_{c3}(\frac{7}{2}^{+},1D)$&  3176$\pm$8 &0 &2&2&1&3&$\frac{7}{2}$\\
 $\Xi_{c1}^{\prime}(\frac{3}{2}^{-},1P)$&2915$\pm$5  &0&1&1&1&1&$\frac{3}{2}$ & $\Xi_{c1}(\frac{1}{2}^{+},1D)$&3281$\pm$8    &2&0&2&1&1&$\frac{1}{2}$  \\
 $\Xi_{c2}^{\prime}(\frac{3}{2}^{-},1P)$&2924$\pm$5  &0&1&1&1&2&$\frac{3}{2}$&  $\Xi_{c1}(\frac{3}{2}^{+},1D)$& 3336$\pm$4   &2&0&2&1&1&$\frac{3}{2}$ \\
 $\Xi_{c2}^{\prime}(\frac{5}{2}^{-},1P)$&2948$\pm$7  &0&1&1&1&2&$\frac{5}{2}$&  $\Xi_{c2}(\frac{3}{2}^{+},1D)$& 3302$\pm$4    &2&0&2&1&2&$\frac{3}{2}$ \\
 $\Xi_{c1}^{\prime}(\frac{1}{2}^{-},1P)$& 3018$\pm$7 &1&0&1&0&1&$\frac{1}{2}$&  $\Xi_{c2}(\frac{5}{2}^{+},1D)$& 3368$\pm$6    &2&0&2&1&2&$\frac{5}{2}$ \\
$\Xi_{c1}^{\prime}(\frac{1}{2}^{-},1P)$&  3038$\pm$7  &1&0&1&0&1&$\frac{3}{2}$&  $\Xi_{c3}(\frac{5}{2}^{+},1D)$&   3362$\pm$6  &2&0&2&1&3&$\frac{5}{2}$ \\
 &&&&&&&  &$\Xi_{c3}(\frac{7}{2}^{+},1D)$& 3393$\pm$9    &2&0&2&1&3&$\frac{7}{2}$\\
 &&&&&&&  &$\Xi_{c2}(\frac{3}{2}^{+},1D)$&  3261$\pm$8   &1&1&2&0&2&$\frac{3}{2}$\\
 &&&&&&&  &$\Xi_{c2}(\frac{5}{2}^{+},1D)$&   3282$\pm$8   &1&1&2&0&2&$\frac{5}{2}$\\
\hline
 \hline
 \end{tabular}
 \end{table*}

\subsection{$\Xi_{c}(2790)$ and $\Xi_{c}(2815)$}
Both these two excited $\Xi_{c}$ baryons were first discovered by the CLEO Collaboration as $J^P=1/2^-$ and $3/2^-$ states~\cite{CLEO:1999msf}. The $\Xi_{c}(2790)$ was observed in the decay channel $\Xi_{c}^{\prime}\pi$~\cite{CLEO:2000ibb}, while $\Xi_{c}(2815)^0$ and
$\Xi_{c}(2815)^+$ were observed in the final states of $\Xi_c^{0(+)}\pi^+\pi^-$ via the intermediate states $\Xi_c(2645)^0$ and $\Xi_c(2645)^+$~\cite{CLEO:1999msf}, respectively. These two states were later be confirmed by the Belle Collaboration in 2016 with resonance parameters as ~\cite{Belle:2016lhy}
\begin{align}
 \label{47}
    \Xi_{c}(2790)^{+}: M &= 2791.6 \pm 0.2 \pm 0.1 \pm 0.4~\mathrm{MeV}, \nonumber \\
    \Gamma &=8.9\pm0.6\pm0.8 ~\mathrm{MeV}; \nonumber \\
    \Xi_{c}(2790)^{0}: M &= 2794.9 \pm 0.3 \pm 0.1 \pm 0.4~\mathrm{MeV}, \nonumber \\
    \Gamma &=10.0\pm0.7\pm0.8 ~\mathrm{MeV}; \nonumber \\
    \Xi_{c}(2815)^{+}: M &= 2816.65 \pm 0.03 \pm 0.03 \pm0.23 ~\mathrm{MeV}, \nonumber \\
    \Gamma &= 2.07\pm0.08\pm0.12~\mathrm{MeV}; \nonumber\\
    \Xi_{c}(2815)^{0}: M &= 2820.20 \pm 0.08 \pm 0.07~\mathrm{MeV}, \nonumber \\
    \Gamma &= 2.54\pm0.18\pm0.17~\mathrm{MeV}. \nonumber
\end{align}
Furthermore, mass and width of $\Xi_{c}(2815)^+$ have been measured by the LHCb Collaboration in $\Xi_c^+\pi^-\pi^+$~\cite{LHCb:2025mge}.

Until now, almost all the theoretical works generally hold that these two particles are $\rho$-mode $P$-wave excitation of $\Xi_c$ with the established quantum numbers. In Ref.~\cite{Bahtiyar:2020uuj}, the authors studied the charmed baryon spectrum within the lattice QCD, and found the masses of $\lambda$-mode excited $\Xi_{c1}(\frac{1}{2}^{-},1P)$ and $\Xi_{c1}(\frac{3}{2}^{-},1P)$ states should be 2770 MeV and 2797 MeV. In Ref.~\cite{Chen:2016phw,Chen:2015kpa}, QCD sum rules have been used and the masses of these two states can been $2.79 \pm 0.15$ GeV and $2.83 \pm 0.15$ GeV. It is worth noting that the mass difference between this doublets in this work is 42 MeV, which is in  agreement with our results. In addition, quark models~\cite{Ebert:2011kk,Roberts:2007ni}, flux tube models~\cite{Chen:2014nyo} and $^{3}P_{0}$ model~\cite{Chen:2007xf} all conducted studies on the spectrum of these two particles, and reached consistent qualitative conclusions.

In our work, mass spectra and decay widths of $\rho$-mode $P$-wave excitation of $\Xi_c$ are shown in Table \ref{4-1} and Table \ref{2790}.

(1) $\mathbf{\Xi_{c}(2790)}$: The theoretical masses of $\Xi_{c1}(\frac{1}{2}^{-})$ is 2783 MeV, extremely close to $\Xi_c(2790)$ and its decay width to $\Xi_{c}^{\prime}\pi$ is 21.13 MeV. Due to the symmetry, we cannot distinguish the $u$ and $d$ quarks. Therefore, in the calculation, $\Xi_c(2790)^0$ and $\Xi_c(2790)^+$ are the same. Namely, the decay width should be the sum of $\Xi_c(2790)^0$ and $\Xi_c(2790)^+$, which is 18.9 MeV in the experiment and very close to our theoretical results. In Ref.~\cite{Chen:2007xf}, the author studied the strong decays of $\Xi_c(2790)$ and $\Xi_c(2815)$ and the result showed that the total width of $\Xi_c(2790)$ decay to $\Xi_{c}^{\prime}\pi$ mode can be 20.2 MeV. Therefore, $\Xi_c(2790)$ can be identified as a $\rho$-mode $P$-wave state.

 (2) $\mathbf{\Xi_{c}(2815)}$: For $\Xi_{c}(2815)$, our results are similar to those obtained in $\Xi_c(2790)$. The theoretical mass of $\Xi_{c1}(\frac{3}{2}^{-})$ is 2830 MeV, close to experimental value.
 There are two decay modes with non-zero numbers, $\Xi_c^{\prime}\pi$ and $\Xi_{c}^{\prime*}\pi$, which is consistent with the conclusion of RPP~\cite{ParticleDataGroup:2024cfk}. The width of the main decay channel is 8.11 MeV and is very close to the sum of $\Xi_c(2815)^0$ and $\Xi_c(2815)^+$. Hence, $\Xi_c(2815)$ can be identified as the $P$-wave excited state and form a doublet $\Xi_{c1}(\frac{1}{2}^{-}, \frac{3}{2}^-)$ with $\Xi_c(2790)$.

\begin{table*}[ht]
\centering
\setlength{\tabcolsep}{13pt}
\caption{The decay widths (in MeV) of $\Xi_{c}(2790)$ and $\Xi_{c}(2815)$ as a candidate for $1P$ wave state, where the states are represented with $J^P$ quantum number, radial and orbital excitation, and mass (in MeV). Here, "$\sim0$" means the corresponding partial decay width is less than 0.01 MeV.. \label{2790}}
\begin{tabular}{*{7}{c}}
\hline \hline
State &  $\Xi_c\pi$ & $\Xi_c^{\prime}\pi$ & $\Xi_{c}^{\prime*}\pi$ & $\Lambda_{c}\bar{K}$ & $\Gamma_{total}$ \\
\hline
$\Xi_{c1}({\frac{1}{2}^{-}},1P,2783)$ & $\sim 0$ &21.13$\pm$1.64  &$\sim 0$  &$\sim 0$  &21.13$\pm$1.64  \\
$\Xi_{c1}({\frac{3}{2}^{-}},1P,2830)$ &  $\sim 0$ & 0.01$\pm$0.0005 &8.11$\pm$0.93  &$\sim 0$  &8.12$\pm$0.93  \\
\hline \hline
\end{tabular}
\end{table*}

 \subsection{$\Xi_{c}(2923)$, $\Xi_{c}(2939)$ and $\Xi_{c}(2965)$}
  In 2020, LHCb Collaboration announced that three $\Xi_c^0$ resonances $\Xi_c(2923)^0$, $\Xi_c(2939)^0$ and $\Xi_c(2965)^0$ were observed in the $\Lambda^+_cK^-$ spectrum, whose masses and widths are shown below as~\cite{LHCb:2020iby}

\begin{align}
    \Xi_{c}(2923)^{0}: M &= 2923.04 \pm 0.25 \pm 0.20 \pm 0.14~\mathrm{MeV}, \nonumber \\
    \Gamma &= 7.1 \pm 0.8 \pm 1.8~\mathrm{MeV}; \nonumber \\
    \Xi_{c}(2939)^{0}: M &= 2939.55 \pm 0.21 \pm 0.17 \pm 0.14~\mathrm{MeV}, \nonumber \\
    \Gamma &= 10.2 \pm 0.8 \pm 1.1~\mathrm{MeV}; \nonumber \\
    \Xi_{c}(2965)^{0}: M &= 2964.88 \pm 0.26 \pm 0.14 \pm 0.14~\mathrm{MeV}, \nonumber \\
    \Gamma &= 14.1 \pm 0.9 \pm 1.3~\mathrm{MeV}. \nonumber
\end{align}
Also, the LHCb Collaboration pointed out that the $\Xi_c(2930)^0$~\cite{Belle:2018yob,BaBar:2007xtc} should be a superposition state due to the overlap of $\Xi_c(2923)^{0}$ and $\Xi_c(2939)^{0}$,  while the $\Xi_c(2970)^0$~\cite{BaBar:2007zjt,Belle:2016lhy,Belle:2006edu} may be different from $\Xi_c(2965)^0$ due to the significant difference in decay width.

The above results have led to heated discussion since the quantum numbers of these resonances are still unknown. In the framework of the QCD sum rule, these three states can be explained as $\lambda$-mode $P$ wave $\Xi_c'$ states with quantum numbers
$\frac{1}{2}^-$ or $\frac{3}{2}^-$~\cite{Yang:2020zjl}. In Ref.~\cite{Wang:2020gkn}, these three $\Xi_c^0$ states and their two body strong decays were evaluated
within a chiral quark model and their results showed that $\Xi_c(2923)^0$ and $\Xi_c(2938)^0$ are most likely to be 1$P$ $\Xi_c'$ states with $J^P=\frac{3}{2}^-$, while the $\Xi_c(2965)^0$ should be 1$P$ $\Xi_c'$ states with $J^P=\frac{5}{2}^-$. After studying the decay widths through $^{3}P_0$ model, Ref.~\cite{Lu:2020ivo} pointed out that $\Xi_c(2923)^0$ and $\Xi_c(2938)^0$ might be an 1$P$ $\Xi_c'$ states but $\Xi_c(2965)^0$ may be a 2$S$ $\Xi_c'$ states. Lattice simulations also give theoretical
investigations of these heavy flavored baryons~\cite{Edwards:2012fx,Padmanath:2013bla,Bahtiyar:2020uuj}. In addition, several theoretical works described these three states as molecular states~\cite{Jimenez-Tejero:2009cyn,Romanets:2012hm,Yu:2018yxl,Nieves:2019jhp,Zhu:2020jke}, including our previous work~\cite{Hu:2020zwc}. Considering  these three states are very close to the 2$S$ and 1$P$ states of $\Xi_c$ and 1$P$ states of $\Xi_c^{\prime}$, all the other 2$S$ and 1$P$ $\Xi_c^{(\prime)}$ configurations have been studied and the mass spectra in addition with decay widths are shown in Table \ref{2923} and Table \ref{2939}.

\begin{table*}[ht]
\centering
\setlength{\tabcolsep}{2.5pt}
\caption{The decay widths (in MeV) of $\Xi_{c}(2923)$, $\Xi_{c}(2930)$, $\Xi_{c}(2939)$, $\Xi_{c}(2965)$ and $\Xi_{c}(2970)$ as the candidates for the $1P$ or $2S$ states of $\Xi_c$, the symbol "$\times$" means this decay can not happen due to threshold, while "$\sim0$" means the corresponding partial decay width is less than 0.01 MeV.\label{2923}}
\begin{tabular}{*{11}{c}}
\hline \hline
State &  $\Xi_c\pi$ & $\Xi_c^{\prime}\pi$ & $\Xi_{c}^{\prime*}\pi$ & $\Lambda_{c}\bar{K}$ & $\Sigma_{c}\bar{K}$ & $\Sigma_{c}^\ast\bar{K}$ & $\Lambda D$ & $\Sigma D$ & $\Gamma_{total}$ \\
\hline
$\Xi_{c}(\frac{1}{2}^{+},2S,2970)$&  $\sim 0$ & 0.74$\pm$0.02 & 27.31$\pm$0.90  & $\sim 0$ & 0.01$\pm$0.05 & $\times$ & $\times$ & $\times$ & 28.06$\pm$0.90 \\
$\Xi_{c0}(\frac{1}{2}^{-},1P,2950)$& 68.71$\pm$1.15 & 65.85$\pm$1.37 & 0.02$\pm$0.0006 &  65.93$\pm$1.63    & $\times$ & $\times$ & $\times$ & $\times$ & 200.51$\pm$2.42 \\
$\Xi_{c1}(\frac{1}{2}^{-},1P,2910)$& 126.69$\pm$2.35& 114.15$\pm$2.75 & 0.01$\pm$0.0004 & 127.54$\pm$3.91 & $\times$ & $\times$ & $\times$ & $\times$ & 368.39$\pm$5.32\\
$\Xi_{c1}(\frac{3}{2}^{-},1P,3057)$&1.75$\pm$0.02 & 5.83$\pm$0.09 & 13.97$\pm$0.27 & 1.65$\pm$0.03 &  0.28$\pm$0.01  &2.99$\pm$0.60 & $\sim 0$ & $\sim 0$ & 26.48$\pm$0.66\\
$\Xi_{c2}(\frac{3}{2}^{-},1P,3037)$&1.51$\pm$0.02 & 4.66$\pm$0.07 & 37.69$\pm$0.78 & 1.41$\pm$0.02 & 0.14$\pm$0.01  & 2.20$\pm$1.83 & $\sim 0$ & $\sim 0$ & 47.61$\pm$1.99  \\
$\Xi_{c2}(\frac{5}{2}^{-},1P,3116)$& 68.94$\pm$0.87&  19.58$\pm$0.29&  40.82$\pm$0.74& 64.38$\pm$0.95 & 2.19$\pm$0.08  &  2.00$\pm$0.13 & $\sim 0$ & $\sim 0$ & 197.91$\pm$1.52\\
\hline \hline
\end{tabular}
\end{table*}

\begin{table*}[ht]
\centering
\setlength{\tabcolsep}{2.0pt}
\caption{The decay widths (in MeV) of $\Xi_{c}(2923)$, $\Xi_{c}(2930)$, $\Xi_{c}(2939)$, $\Xi_{c}(2965)$ and $\Xi_{c}(2970)$ as the candidate for the $1P$ or $2S$ state of $\Xi_c^{\prime}$, where the states are represented with $J^P$ quantum numbers, radial and orbital excitations, and masses (in MeV). Here, the symbol "$\times$" means this decay can not happen due to threshold, and "$\sim0$" means the corresponding partial decay width is less than 0.01 MeV. \label{2939}}
\begin{tabular}{*{10}{c}}
\hline \hline
State & $\Xi_c\pi$ & $\Xi_c^{\prime}\pi$ & $\Xi_{c}^{\prime*}\pi$ & $\Lambda_{c}\bar{K}$ & $\Sigma_{c}\bar{K}$ & $\Sigma_{c}^\ast\bar{K}$ & $\Lambda D$ & $\Sigma D$ & $\Gamma_{total}$ \\
\hline
$\Xi_{c}^{\prime}(\frac{1}{2}^{+},2S,3054)$ & 3.58$\pm$0.05 & 1.96$\pm$0.03 & 22.20$\pm$0.50 & 57.33$\pm$1.05 & 2.80$\pm$0.17 & 0.21$\pm$0.06 & 36.11$\pm$3.21 & 2.22$\pm$0.31 & 126.41$\pm$3.44\\
$\Xi_{c}^{*}(\frac{3}{2}^{+},2S,3101)$ & 9.19$\pm$0.12 & 1.35$\pm$0.02 & 35.42$\pm$0.68 & 72.33$\pm$1.12 & 2.90$\pm$0.11 & 2.02$\pm$0.16 & 44.33$\pm$2.33 & 4.71$\pm$0.33 & 172.25$\pm$2.70\\
$\Xi_{c0}^{\prime}(\frac{1}{2}^{-},1P,2874)$ & 54.73$\pm$1.19 & 31.05$\pm$0.95 & $\sim0$ & 54.04$\pm$2.30 & $\times$ & $\times$ & $\times$ & $\times$ & 139.82$\pm$2.75\\
 $\Xi_{c1}^{\prime}(\frac{1}{2}^{-},1P,2908)$& 23.72$\pm$0.46 & 18.47$\pm$0.47 & $\sim0$ & 23.02$\pm$0.74 & $\times$ & $\times$ & $\times$ & $\times$ & 65.21$\pm$0.99\\
$\Xi_{c1}^{\prime}(\frac{3}{2}^{-},1P,2915)$ & 0.13$\pm$0.003 & 0.26$\pm$0.007 & 14.47$\pm$0.57 & 0.10$\pm$0.003 & $\times$ & $\times$ & $\times$ & $\times$ & 14.96$\pm$0.57\\
$\Xi_{c2}^{\prime}(\frac{3}{2}^{-},1P,2924)$&0.17$\pm$0.003 & 0.33$\pm$0.008 & 23.52$\pm$0.87 & 0.12$\pm$0.004 & $\times$ & $\times$ & $\times$ & $\times$ & 24.14$\pm$0.87\\
$\Xi_{c2}^{\prime}(\frac{5}{2}^{-},1P,2948)$ & 13.58$\pm$0.26& 1.75$\pm$0.04 & 2.14$\pm$0.08 & 12.54$\pm$0.36 & $\times$ & $\times$ & $\times$ & $\times$ & 30.01$\pm$0.45\\
$\Xi_{c1}^{\prime}(\frac{1}{2}^{-},1P,3018)$ & $\sim0$ & 98.08$\pm$1.84 & $\sim0$ & $\sim0$ & 32.53$\pm$3.30 & $\times$ & 0.04$\pm$0.007 & $\sim0$ & 130.65$\pm$3.77\\
$\Xi_{c1}^{\prime}(\frac{3}{2}^{-},1P,3038)$ & $\sim0$ & $\sim0$ & 89.03$\pm$2.03 & $\sim0$ & $\sim0$ & $\times$ &$\sim0$ & $\sim0$ & 89.03$\pm$2.03\\
\hline \hline
\end{tabular}
\end{table*}

(1)~$\mathbf{\Xi_{c}(2923)}$:
According to Table \ref{2923} and Table \ref{2939}, the mass of $\Xi_{c2}^{\prime}(\frac{3}{2}^{-})$ is almost the same as $\Xi_{c}(2923)$. However, its decay width to $\Lambda_c \overline{K}$ is very small, which is contrary to the experimental results. Thus $\Xi_{c2}^{\prime}(\frac{3}{2}^{-})$ may be excluded. Similarly, $\Xi_{c1}^{\prime}(\frac{3}{2}^{-})$, $\Xi_{c0}(\frac{1}{2}^{-})$ and $\Xi_{c1}(\frac{1}{2}^{-})$ should also be ruled out. While for $\Xi_{c1}(\frac{3}{2}^{-})$, $\Xi_{c2}(\frac{3}{2}^{-})$, $\Xi_{c2}(\frac{5}{2}^{-})$, $\Xi_{c1}^{\prime}(\frac{1}{2}^{-})$ and $\Xi_{c1}^{\prime}(\frac{3}{2}^{-})$, these states also should be excluded due to their too large masses. Then, since the mass of $\Xi_{c1}^{\prime}(\frac{1}{2}^{-})$ is 2908 MeV and its partial decay width to $\Lambda_c \overline{K}$ final state is 21.6 MeV, and considering that our work does not account for unquenched effects, the theoretical results will be higher than the experimental values. Therefore, we are inclined to consider $\Xi_{c1}^{\prime}({\frac{1}{2}^{-}})$ as a candidate for $\Xi_{c}(2923)$.

(2)~$\mathbf{\Xi_{c}(2939)}$:
Since its mass and decay width are very close to those of $\Xi_{c}(2923)$, we obtain similar circumstances during our analysis. After excluding several states whose masses and widths are significantly different from the experimental values, we found $\Xi_{c2}^{\prime}(\frac{5}{2}^{-})$ can be a good candidate for $\Xi_{c}(2939)$ because our theoretical mass and width are both extremely close to experimental results.

(3)~$\mathbf{\Xi_{c}(2965)}$:
The results shown in Table \ref{2923} and Table \ref{2939} indicate that  there are two states that can be candidates for $\Xi_{c}(2965)$, $\Xi_{c}(\frac{1}{2}^{+},2S)$ and $\Xi_{c0}(\frac{1}{2}^{-},1P)$ from the perspective of mass. However, the decay of $\Xi_{c}(\frac{1}{2}^{+},2S)$ to $\Lambda_c \overline{K}$ is strictly forbidden. Therefore, $\Xi_{c}(2965)$ are tend to be identified as  a $\rho$-mode $P$-wave excitation of $\Xi_{c}$ with quantum number of $J^P=\frac{1}{2}^-$.

It is important to highlight here that we also discovered a $\Xi_{c0}^{\prime}(\frac{1}{2}^{-})$ state and its mass is 2874 MeV, which is very close to $\Xi_{c}(2882)$ observed by LHCb Collaboration in 2023~\cite{LHCb:2022vns}. We hope that subsequent experiments will be able to find this state and thereby confirm our conclusion, although the status of $\Xi_{c}(2882)$ is not so clear at present.

 \subsection{$\Xi_{c}(2970)$}

 The $\Xi_{c}(2970)$ was observed by the Belle Collaboration in the $\Lambda_{c}^{+}K^{-}\pi^{+}$ and $\Lambda_{c}^{+}K_S^{0}\pi^{-}$ invariant mass spectra in 2006~\cite{Belle:2006edu} and it was confirmed later by Belle in $\Xi_c^*\pi~$channel~\cite{Belle:2008yxs}. In 2020, the Belle experiment determined its quantum number to be $\frac{1}{2}^{+}$ by measuring the angular distribution of $\Xi_{c}(2970)^{+}\rightarrow \Xi_{c}(2645)^{0}\pi^{+} \rightarrow \Xi_{c}^{+}\pi^{-}\pi^{+}$ \cite{Belle:2020tom}, which is the first determination of the spin and parity of a charmed-strange baryon. Furthermore, the decay branching ratio of $\Xi_{c}(2970)$ also presented by Belle as
\begin{eqnarray}
&&\frac{\Gamma_{\Xi_{c}(2970)^{+}\rightarrow \Xi_{c}(2645)^{0} \pi^{+}}}{\Gamma_{\Xi_{c}(2970)^{+}\rightarrow \Xi_{c}^{\prime^{0}}\pi^{+}}} \\\nonumber&&\qquad= 1.67 \pm 0.29 (\text{stat})^{+0.15}_{-0.09} (\text{sys}) \pm 0.25 (\text{IS}).
\end{eqnarray}
where IS refers to the uncertainty due to possible isospin symmetry-breaking effects. In addition, this $R$-value favors the spin of the light-quark degrees of freedom $S_{\rho}=0$.
 Here, we take the decay width of $\Xi_{c}(2970)^{+}$ from RPP as $20.9^{+2.4}_{-3.5}~\mathrm{MeV}$ \cite{ParticleDataGroup:2024cfk}.

Since the quantum number $J^P=\frac{1}{2}^{+}$ of $\Xi_{c}(2970)$ is already clear, then it can only be a $2S$ state or $1D$ state. Several quark-model-based calculations are consistent with the experimental results. In Ref.~\cite{Ebert:2011kk}, mass of the $\Xi_{c}(2S)$ were calculated to be 2983 MeV in the heavy-quark–light-diquark picture in the framework of the QCD-motivated relativistic quark model, suggesting that $\Xi_{c}(2970)$ can be $2S$ state of $\Xi_{c}$. In Ref.~\cite{Chen:2014nyo}, $\Xi_{c}(2970)$ was studied by the relativistic flux tube model and assigned to the first radial excitations with $J^P=\frac{1}{2}^+$. In Refs.~\cite{Shah:2016mig,Migura:2006ep}, the hypercentral description of three body system and relativistically covariant quark model
based on the Bethe-Salpeter-equation are both used to investigated the $\Xi_{c}(2970)$ and the showed that it is indeed possible to be a $2S$ state. Furthermore, $^{3}P_{0}$ model is also used to study the decay widths and certain ratio of branching fractions of $\Xi_{c}(2970)$ and its results are compatible with the experimental data~\cite{Zhao:2020tpf}. 

(1)~$\mathbf{\Xi_{c}(2970)}$:
From the perspective of the mass spectrum, according to Table~\ref{4-1} and Table~\ref{4-2} , the mass of $\Xi_{c}(\frac{1}{2},2S)$ at $2970~\mathrm{MeV}$ is the same as the $\Xi_{c}(2970)$ while other excitations with positive parity are all over 3000 MeV, which must be excluded. The decay of $\Xi_{c}(2970)$ into $\Lambda_c \overline{K}$ or $\Xi_c\pi$ has never been seen in experiments, which is consistent with our results. The latest experimental data shows that the width of $\Xi_{c}(2970)$ is $\Gamma=20.9^{+2.4}_{-3.5}$~MeV~\cite{ParticleDataGroup:2024cfk}, close to our results. Considering the mass spectrum and decay width, we tend to interpret $\Xi_{c}(2970)$ as the $\Xi_{c}(\frac{1}{2}^{+},2S)$ state.

\subsection{$\Xi_{c}(3055)$ and $\Xi_{c}(3080)$}
 The $\Xi_{c}(3080)$ was also discovered by the Belle Collaboration in the $\Lambda_c^+K^-\pi^+$ and $\Lambda_{c}^{+}K_{s}^{0}\pi^{-}$ invariant mass spectra in 2006 \cite{Belle:2006edu}. Later, it was discovered in $\Sigma_{c}(2455)^{++}K^{-}$ and $\Sigma_{c}(2520)^{++}K^{-}$ decay channels by BABAR Collaboration in 2008~\cite{BaBar:2007zjt}. In 2016, the Belle Collaboration review the $\Xi_{c}(3055)$ and $\Xi_{c}(3080)$ in the $\Lambda D$ decay mode and provided the masses and decay widths of all the states as \cite{Belle:2016tai}

\begin{eqnarray}
\Xi_{c}(3055)^{0}: M &=& 3059.0 \pm 0.5 \pm 0.6~\mathrm{MeV}, \nonumber \\
\Gamma &=& 6.4 \pm 2.1 \pm 1.1~\mathrm{MeV}; \nonumber \\
\Xi_{c}(3055)^{+}: M &=& 3055.8 \pm 0.4 \pm 0.2~\mathrm{MeV}, \nonumber \\
\Gamma &=& 7.0 \pm 1.2 \pm 1.5~\mathrm{MeV}; \nonumber \\
\Xi_{c}(3080)^{+}: M &=& 3079.6 \pm 0.4 \pm 0.1~\mathrm{MeV}, \nonumber \\
\Gamma &=& 3.0 \pm 0.7 \pm 0.4~\mathrm{MeV}.
\end{eqnarray}

It is worth noting that the spin and parity of $\Xi_{c}(3055)$ have been determined by the LHCb Collaboration to be $J^P = 3/2^+$, which suggests it to be the first identified $\lambda$-mode $D$ wave excitation of $\Xi_c$ \cite{LHCb:2024eyx}. And for 
$\Xi_{c}(3080)^{+}$, its width and mass have also been measured in the invariant mass spectrum of $\Xi_c^+\pi^-\pi^+$ by the LHCb Collaboration as ~\cite{LHCb:2025mge}
\begin{eqnarray}
\Xi_{c}(3080)^{+}: M &=& 3076.8 \pm 0.7 \pm 1.3 
 \pm0.2~\mathrm{MeV}, \nonumber \\
\Gamma &=& 6.8 \pm 2.3 \pm 0.9~\mathrm{MeV}.
\end{eqnarray}

Due to the specific quantum numbers of $\Xi_{c}(3080)$ have not been fully determined, various works with different models have been used to study its quantum number.
\iffalse For $\Xi_{c}(3055)$, before the experimental measurement, Refs.~\cite{Valcarce:2008dr,Guo:2008he} already suggested it can be perfectly described as a $D$ wave excited $\Xi_{c}$ state with $J^P=\frac{5}{2}^+$. In Refs~\cite{Ebert:2011kk,Chen:2015kpa,Chen:2016phw,Chen:2014nyo,Liu:2012sj},  $\Xi_{c}(3055)$ can be a D wave state with $J^P=\frac{5}{2}^+$ or $\frac{7}{2}^+$.
For $\Xi_{c}(3080)$, \fi For example, Ref.~\cite{Valcarce:2008dr} considers it to be an excitation of $\Xi_{c}$ with $J^P=\frac{3}{2}^+$.  In Refs.~\cite{Guo:2008he,Cheng:2006dk,Cheng:2015naa,Ebert:2011kk,Chen:2014nyo}, the authors 
regarded $\Xi_{c}(3080)$ as a $D$ wave state with $J^P=\frac{5}{2}^+$. In Refs.~\cite{Chen:2015kpa,Chen:2016phw}, the authors consider $\Xi_{c}(3080)$ to be a $\Xi_{c}^{\prime}(1P)$ state with $J^P=\frac{5}{2}^-$or a $\Xi_{c}(1D)$ state with $J^P=\frac{5}{2}^+$, while Ref.~\cite{Liu:2012sj} suggests it is a $\Xi_{c}(2S)$ state with $J^P=\frac{1}{2}^+$. In Ref.\cite{Wang:2023wii,Wang:2017vtv}, $\Xi_{c}(3080)$ have been investigated as a $D$-wave baryon state in the framework of QCD sum rule.

(1)~$\mathbf{\Xi_{c}(3055)}$: According to Table \ref{4-1} and Table \ref{4-2}, both $\Xi_{c}^{\prime}(\frac{1}{2}^{+},2S)$ and $\Xi_{c2}(\frac{3}{2}^{+},1D)$ are close to $\Xi_{c}(3055)$ with positive parity. However, the decay width of $\Xi_{c}^{\prime}(\frac{1}{2}^{+},2S)$ is much higher than experimental value and it cannot be a good candidate of $\Xi_{c}(3055)$. We further calculated the decay width of $\Xi_{c2}(\frac{3}{2}^{+},1D)$ by $^{3}P_{0}$ model and the results are shown in Table \ref{3055}. The total decay width is 13.83 MeV, very close to the the sum of the central values of decay with of $\Xi_{c}(3055)^0$ and $\Xi_{c}(3055)^+$. Therefore, $\Xi_{c}(3055)$ can be interpreted as a $\lambda$ mode $1D$ wave $\Xi_c$ baryons, which is consistent with the experimental conclusion~\cite{LHCb:2024eyx}.

(2)~$\mathbf{\Xi_{c}(3080)}$:
Based on our research on $\Xi_{c}(3055)$, we found that the mass of $\Xi_{c2}(\frac{5}{2}^{+},1D)$ is 3078 MeV, which is almost the same as that of $\Xi_{c}(3080)$. For $\Xi_{c}(3080)^0$, its total decay width is $5.6\pm2.2$ MeV~\cite{ParticleDataGroup:2024cfk}, and the sum of this width and that of $\Xi_{c}(3080)^+$ is around 8.6 MeV, which is almost the same as our calculation as 8.86 MeV. Therefore, we consider $\Xi_{c2}(\frac{5}{2}^{+},1D)$ is a good arrangement for $\Xi_{c}(3080)$, which can form a doublet as $\Xi_{c2}(\frac{3}{2}^+,\frac{5}{2}^+)$ with the $\Xi_{c}(3055)$.

\begin{table*}[ht]

\centering
\setlength{\tabcolsep}{3pt}
\caption{The decay widths (in MeV) of $\Xi_{c}(3055)$ and $\Xi_{c}(3080)$ as candidates for $1D$ states, where the states are represented with $J^P$ quantum number, radial and orbital excitation, and mass (in MeV). Here, "$\sim0$" means the corresponding partial decay width is less than 0.01 MeV. \label{3055}}
\begin{tabular}{cccccccccccccc}
\hline \hline
State &$\Xi_c\pi$~   &$\Xi_c^{\prime}\pi$~ &$\Xi_{c}^{*}\pi$ &$\Lambda_{c}\bar{K}$ &$\Sigma_{c}\bar{K}$ &$\Sigma_{c}^{*}\bar{K}$ &$\Lambda D$ &  $\Sigma D$  &$\Gamma_{total}$ \\
\hline
$\Xi_{c2}(\frac{3}{2}^{+},1D,3043)$&  $\sim 0$ & 9.10$\pm$0.17 & 2.00$\pm$0.05 & $\sim 0$ & 1.22$\pm$0.09 & 0.01$\pm$0.007 & 0.70$\pm$0.07 & 0.80$\pm$0.15 & 13.83$\pm$0.25\\
$\Xi_{c2}(\frac{5}{2}^{+},1D,3078)$& $\sim 0$ & 2.36$\pm$0.04 & 5.32$\pm$0.11 & $\sim 0$ & 0.09$\pm$0.004 & 0.57$\pm$0.07 & 0.24$\pm$0.02 & 0.27$\pm$0.03 & 8.85$\pm$0.14 \\
\hline \hline
\end{tabular}
\end{table*}

\subsection{$\Sigma$-type $D$-wave $\Xi_{c}^{\prime}$ states}
We also studied the $\lambda$-mode $D$ wave excitation of $\Xi_{c}^{\prime}$, and the masses and corresponding decay width are shown in Table \ref{3123}, where the masses of these states are very close to $\Xi_c(3123)$. In addition, their decay widths are all around several MeV. For the $\Xi_{c}(3123)$, it was observed by the BABAR experiment in the mass spectra of $\Lambda_{c}^{+}K^{-}\pi^{+}$ in 2008~\cite{BaBar:2007zjt}, and its mass and decay width were reported as follows
\begin{eqnarray}
\label{m3123}
\Xi_{c}(3123)^{+}: M &=& 3122.9 \pm 1.3 \pm 0.3~\mathrm{MeV}, \nonumber \\
\Gamma &=& 4.4 \pm 3.4 \pm 1.7~\mathrm{MeV}.
\end{eqnarray}
Then, in 2014, the Belle experiment attempted to find the signal of $\Xi_{c}(3123)$ in the final state of $\Sigma_{c}(2520)^{++}K^{-}$ but did not observe it \cite{Belle:2013htj}. On the PDG, the statistical significance of $\Xi_{c}(3123)$ is $3.6\sigma$, indicating that the experimental information is still insufficient to confirm it as a new particle state.
While on the theoretical side, in Ref.~\cite{Chen:2014nyo}, the authors suggested it as a $\Xi_{c}(2P)$ state with $J^P=\frac{1}{2}^-$ or $\frac{3}{2}^{-}$.

However, based on our calculation on the mass spectra and partial decay width, it is highly likely that $\Xi_{c}(3123)$ is a $\Sigma$-type $D$-wave excitation of $\Xi_{c}^{\prime}$, although its spin-parity is unknown. From our results in Table \ref{3123}, the decay widths of all $D$-wave excitation of $\Xi_{c}^{\prime}$ to the $\Sigma_{c}\bar{K}$ are very small. For $\Xi_{c1}^{\prime}(\frac{1}{2}^{+})$ and $\Xi_{c3}^{\prime}(\frac{7}{2}^{+})$, the main decay channels for both of them are $\Lambda D$ and $\Sigma D$. For $\Xi_{c1}^{\prime}(\frac{3}{2}^{+}$), its main decay channels are $\Xi_c\pi$ and $\Xi_{c}^{*}\pi$. While for $\Xi_{c2}^{\prime}(\frac{5}{2}^{+})$ and $\Xi_{c2}^{\prime}(\frac{3}{2}^{+})$, their main decay channel is $\Xi_{c}^{*}\pi$. Finally, for $\Xi_{c3}^{\prime}(\frac{5}{2}^{+})$, its main decay channels are $\Xi_{c}^{*}\pi$, $\Lambda_{c}\bar{K}$ and $\Sigma_{c}^{*}\bar{K}$. Thus, although their mass positions are close to each other, their dominant decay modes are very different. For $\Xi_{c1}^{\prime}(\frac{1}{2}^{+})$, which is interpreted as $\Xi_{c}(3123)$ in this work, the experiment can attempt to find it in the $\Lambda D$ and $\Sigma D$ decay mode. However it is not easily detectable in the $\Sigma_{c}^{*}\bar{K}$ decay mode due to the corresponding decay width we calculated is 0, which is consistent with the conclusions in Ref.~\cite{Belle:2013htj}. Similarly, for other $\Sigma$-type $D$-wave $\Xi_{c}^{\prime}$ states, the experiment can also search for them using the same method, namely utilizing the dominant decay modes. Therefore, we hope that our results will help the experiment to further identify the nature of $\Xi_{c}(3123)$, and to discover more $D$-wave excitation of $\Xi_{c}^{\prime}$ in the future as a verification of our model.

\begin{table*}[ht]

\centering
\setlength{\tabcolsep}{2.0pt}
\caption{The decay widths (in MeV) of  $\lambda$-mode $D$ wave excitation of $\Xi_{c}^{\prime}$, where the states are represented with $J^P$ quantum numbers, radial and orbital excitations, and masses (in MeV). Here, "$\sim0$" means the corresponding partial decay width is less than 0.01 MeV. \label{3123}}
\begin{tabular}{ccccccccccccc}
\hline \hline
State      &$\Xi_c\pi$~   &$\Xi_c^{\prime}\pi$~ &$\Xi_{c}^{*}\pi$ &$\Lambda_{c}\bar{K}$ &$\Sigma_{c}\bar{K}$ &$\Sigma_{c}^{*}\bar{K}$ &$\Lambda D$ &  $\Sigma D$  &$\Gamma_{total}$ \\
\hline
$\Xi_{c1}^{\prime}(\frac{1}{2}^{+},1D,3136)$ & $\sim 0$ & 0.1$\pm$0.001 & $\sim 0$ & $\sim 0$ & $\sim 0$ & $\sim 0$ & 2.46$\pm$0.10 & 1.32$\pm$0.07 & 3.78$\pm$0.12\\
$\Xi_{c1}^{\prime}(\frac{3}{2}^{+},1D,3143)$ & 1.12$\pm$0.01 & 0.45$\pm$0.005 & 2.73$\pm$0.04 & 0.16$\pm$0.002 & 0.15$\pm$0.004 & 0.60$\pm$0.02 & 0.30$\pm$0.01 & 0.03$\pm$0.001 & 5.54$\pm$0.05 \\
$\Xi_{c2}^{\prime}(\frac{5}{2}^{+},1D,3153)$ & 0.05$\pm$0.0005 & 0.22$\pm$0.003 & 2.03$\pm$0.03 & 0.05$\pm$0.0006 & 0.02$\pm$0.0005 & 0.35$\pm$0.01 & 0.60$\pm$0.02 & 0.04$\pm$0.002 & 3.36$\pm$0.04 \\
$\Xi_{c3}^{\prime}(\frac{5}{2}^{+},1D,3170)$ & 0.05$\pm$0.0005 & 0.23$\pm$0.003 & 2.09$\pm$0.03 & 1.10$\pm$0.01 & 0.50$\pm$0.01 & 1.85$\pm$0.06 & 0.93$\pm$0.03 & 0.05$\pm$0.002 & 6.80$\pm$0.07 \\
$\Xi_{c2}^{\prime}(\frac{3}{2}^{+},1D,3172)$ & 0.72$\pm$0.007 & 0.40$\pm$0.004 & 2.64$\pm$0.03 & 0.01$\pm$0.0001 & 0.15$\pm$0.003 & 0.70$\pm$0.02 & 0.27$\pm$0.01 & 0.02$\pm$0.001 & 5.91$\pm$0.04 \\
$\Xi_{c3}^{\prime}(\frac{7}{2}^{+},1D,3176)$ & $\sim 0$ & $\sim 0$ & $\sim 0$ & $\sim 0$ & $\sim 0$ & $\sim 0$ & 0.92$\pm$0.03 & 0.06$\pm$0.002 & 0.98$\pm$0.03 \\
\hline \hline
\end{tabular}
\end{table*}

\section{summary}
Due to its special constituent quark components, $\Xi_c^{(\prime)}$ family becomes a great platform to study the heavy quark symmetry and chiral symmetry. Especially, the explicit breaking of chiral symmetry makes exchanges of massive goldstone bosons between light quarks possible, and provides an excellent environment for the examination of the chiral constituent quark model.

Until now, there already exist so many $\Xi_c^{(\prime)}$ in RPP~\cite{ParticleDataGroup:2024cfk}. However, the properties of most of them still remain unclear, although there already exist many studies, the classification of these states is still confused \cite{Bahtiyar:2020uuj,Chen:2016phw,Chen:2015kpa,Ebert:2011kk,Roberts:2007ni,Chen:2014nyo}.

However, things may change with the measurements carried out by the LHCb Collaboration, which determines that $\Xi_c(3055)$ is a $\bar{3}_F~\lambda$-mode $1D$ state with quantum number $\frac{3}{2}^+$. In our view, this result can allow $\Xi_c(3055)$ serve as a scaling point, which can be used to restrict the model parameters for further studies on the $\Xi_c^{(\prime)}$ family. Thus, inspired by this, in this work, we investigate the 2$S$, 1$P$ and $1D$ states of $\Xi_{c}^{(\prime)}$ baryons in the framework of the chiral quark model and $^{3}P_{0}$ decay mechanism with the help of Gaussian expansion method. By comparing the results of the mass  spectra and decay widths with the experimental values, we give interpretations for the quantum numbers of these states as follows,

(1) $\Xi_{c}(2790)$ and $\Xi_{c}(2815)$ are probably $\lambda$-mode 1$P$ states and form a doublet $\Xi_{c1}(\frac{1}{2}^-,\frac{3}{2}^-)$. The decay widths of the dominant decay channels of our calculations are very close to experimental values.

(2) $\Xi_{c}(2923)$ and $\Xi_{c}(2939)$ can be $\Xi_{c1}^{\prime}({\frac{1}{2}^{-}},1P)$ and $\Xi_{c2}^{\prime}({\frac{5}{2}^{-}},1P)$, respectively. $\Xi_{c}(2965)$ can be a $\rho$-mode $P$-wave excited state with quantum number $J^P=\frac{1}{2}^-$. In addition, another $P$-wave excited state $\Xi_{c0}^{\prime}(\frac{1}{2}^{-})$ with a mass of 2874 MeV is obtained, and it can be a good candidate for $\Xi_c(2882)$.

(3) $\Xi_{c}(2970)$ can be interpreted as $\Xi_{c}(\frac{1}{2}^{+},2S)$, which is the only possibility in our model, in terms of both mass and decay width.

(4) $\Xi_{c}(3055)$ can be arranged as $\Xi_{c2}(\frac{3}{2}^{+},1D)$, which is the same as the experimental result~\cite{LHCb:2024eyx}, and can form a doublet $\Xi_{c2}(\frac{3}{2}^+,\frac{5}{2}^+)$ with $\Xi_{c}(3080)$, as is the case with $\Xi_{c}(2790)$ and $\Xi_{c}(2815)$.

(5) $\Xi_{c}(3123)$ might be arranged as a $\Sigma$-type $D$-wave $\Xi_{c1}^\prime(\frac{1}{2}^+,1D)$ state, but to determine its detailed quantum number, experimental measurements on the $\Lambda D$ and $\Sigma D$ (its dominant decay modes by our calculations) is critical.

Thus, we hope future experiments can carry  out  relevant measurements to verify our conclusions, which will not only help in further determinations on the properties of the currently unclear $\Xi_c^{(\prime)}$ states, but also will deepen our understandings on the symmetry, strong interaction, and few-body system.

\if
In this work, we only use one unified set of parameters to investigate the mass spectra and decay widths, which means the unquenched effect is not taken into account. Although our results are slightly different from the experimental values, but they are qualitatively consistent. However, we cannot jump to conclusions in this paper. A mixed system of three-quark baryon and pentaquark cannot be ignored. Therefore, the study of $\Xi_c$ in the framework of the unquenched quark model, including the higher Fock components with more future experimental and theoretical data is our future work.
\fi

\section*{Acknowledgments}
This work is supported partly by the National Natural Science Foundation of China under Grant Nos. 12305087, 12205249, the Start-up Funds of Nanjing Normal University under Grant No.~184080H201B20, the Funding for School-Level Research Projects of Yancheng Institute of Technology under Grant No. xjr2022039, the General Project of Natural Science Foundation of colleges and universities of Jiangsu Province under Grant No. 24KJB140001, and the Qinglan Project of Jiangsu Province.

\section*{Appendix}
\subsection{Calculation on the errors of mass spectrum}

During the fit on the mesons and baryons given in Table~\ref{4-3}, we can get not only the values of parameters, but also the errors of them. Thus, we can use these values to calculate the errors of mass spectrum.

We start from the definition of energy in quantum mechanics, i.e., if we already get the wave function $\psi$ of the system that obeys the Hamiltonian $H(\lambda_s)$, with $\lambda_s$ being the parameters in the Hamiltonian, the energy of the system can be calculated as
\begin{eqnarray}
    E(\lambda_s)=\langle \psi(\lambda_s) | H(\lambda_s) | \psi(\lambda_s) \rangle,
\end{eqnarray}
where $\psi(\lambda_s)$ can be expanded by a series of basis functions $\psi_i$ with $\lambda_s$-dependent coefficients as
\begin{eqnarray}
    \psi(\lambda_s)=\sum_i C_i(\lambda_s)\psi_i.
\end{eqnarray}

Then, following the thinking of perturbation theory, we think that the leading order errors may be sufficiently enough. Thus, we can estimate the error of energy as
\begin{eqnarray}
    \delta E = \sum_{s,i,j} C_i(\lambda_s) C_j^\ast(\lambda_s) \langle \psi_j | \frac{\partial}{\partial\lambda_s}H(\lambda_s) | \psi_i \rangle \delta \lambda_s.
\end{eqnarray}
Here, we want to emphasize that to get the above equation, we introduce an assumption that expansion coefficients $C_i(\lambda_s)$ changes little when the parameters vary, i.e., $\frac{\partial}{\partial\lambda_s}C_i(\lambda_s)\approx0$ when $\delta \lambda_s$ are small. However, according to our crosscheck, the error obtained with this method is not the same but very close to that from another strategy, where we varies the model parameters within their permitted ranges to make a dense calculation on the spectra, and then extract the minimum and maximum values to obtain the error.

Thus, when coming to practical calculation, the masses of quarks, the contribution from meson exchange, and the $a_s$, $r_0$, $r_g$ in the non-central potential are fixed as in Table~\ref{4-0}. Thus, the error of the mass spectrum just come from the $\alpha_s$ in OGE potential, in addition with the $a_c$, and $\Delta$ in confinement, then we will have
\begin{eqnarray}
      \delta V_{CON}^C&=& \boldsymbol{\lambda^c_i} \cdot \boldsymbol{\lambda^c_j}\left[-r_{ij}\delta a_{c}-\delta\Delta\right],\\
      \delta V_{CON}^{SO}&=&-\boldsymbol{\lambda^c_i} \cdot \boldsymbol{\lambda^c_j}\frac{r_{ij}\delta a_c}{4m_i^2m_j^2}\left[((m_i^2+m_j^2)(1-2a_s)\right.\nonumber\\
      &&+4m_im_j(1-a_s))(\mathbf S_+\cdot\mathbf L)\nonumber\\
      &&\left.+(m_j^2-m_i^2)(1-2a_s)(\mathbf S_{-}\cdot\mathbf L)\right],\\
      \delta V_{OGE}^C &=& \frac{\delta\alpha_s}{4} \boldsymbol{\lambda}_i^c \cdot \boldsymbol{\lambda}_j^c
   \left[ \frac{1}{r_{ij}}-\frac{\boldsymbol{\sigma}_i\cdot \boldsymbol{\sigma}_j}{6m_im_j}  \frac{e^{-r_{ij}/r_0(\mu)}}{r_{ij}r^2_0(\mu)}\right],\\
      \delta V_{OGE}^T&=&-\frac{1}{16} \frac{\delta\alpha_{s}}{m_im_j} \boldsymbol{\lambda^c_i} \cdot \boldsymbol{\lambda^c_j} \left[ \frac{1}{r_{ij}^3}- \frac{e^{-r_{ij}/r_g(\mu)}}{r_{ij}}\right.\nonumber\\
      &&\left.\times\left(\frac{1}{r_{ij}^2} +\frac{1}{3r_{g}^2(\mu)}+\frac{1}{r_{ij}r_g(\mu)}\right) \right]S_{ij},\\
      \delta V_{OGE}^{SO}&=&-\frac{1}{16} \frac{\delta\alpha_{s}}{m_i^2m_j^2} \boldsymbol{\lambda^c_i} \cdot \boldsymbol{\lambda^c_j} \left[ \frac{1}{r_{ij}^3}- \frac{e^{-r_{ij}/r_g(\mu)}}{r_{ij}^3}\right.\nonumber\\
      &&\left.\times\left(1+\frac{r_{ij}}{r_g(\mu)} \right)\right] \left[((m_i+m_j)^2 +2m_im_j)\right.\nonumber\\
      &&\left.\times(\mathbf{S}_+ \cdot \mathbf{L} )+(m_j^2-m_i^2)(\mathbf{S}_- \cdot \mathbf{L} ) \right].
 \end{eqnarray}
Thus, the Hamiltonian dependence on the parameters is
\begin{eqnarray}
    \sum_s \frac{\partial}{\partial\lambda_s}H(\lambda_s)\delta\lambda_s=\delta V_{CON} + \delta V_{OGE}.
\end{eqnarray}
Finally, to avoid the uncertainties of the signs of errors of parameters, we calculate the errors as
\begin{eqnarray}
    \delta E = \sqrt{\sum_{s,i,j} \left(C_i C_j^\ast \langle \psi_j | \frac{\partial}{\partial\lambda_s}H(\lambda_s) | \psi_i \rangle\right)^2 \left(\delta \lambda_s\right)^2}.
\end{eqnarray}

\subsection{Calculation on the errors of two-body decays}

For the two-body partial decay width of the process $A \to B+C$, we have
\begin{eqnarray}
    \Gamma_{A \to B+C} \equiv \Gamma_{2,i} \sim \frac{p(M_A,M_B,M_C)}{M_A^2} |\mathcal{M}_{A \to B+C}|^2,
\end{eqnarray}
where $p$ is the module of the three momentum of final particles in center-of-mass system, and $\mathcal{M}_{A \to B+C}$ is the transition amplitude.

Ordinary, in quark model, $\mathcal{M}_{A \to B+C}$ can be symbolic expressed as
\begin{eqnarray}
    \mathcal{M}_{A\to B+C} \sim \langle \psi_B \psi_C |\hat{T} |\psi_A\rangle,
\end{eqnarray}
where $\psi_i$ is the wave function of particle $i$, and $\hat{T}$ is the transition operator. Since in quark pair creation model, the parameters in $\hat{T}$ are fixed previously, thus, still following the assumption that $C_i(\lambda_s)$ remain unchanged if $\delta\lambda_s$ is small, we can intermediately infer that $(\mathcal{M}_{A\to B+C})$ will be a constant in error calculation. Thus, we can approximately estimate the error of the two-body decay as
\begin{eqnarray}
    (\delta\Gamma_{2,i})^2 &=& \left(\frac{\partial\Gamma_{2,i}}{\partial M_A}\right)^2 \left(\delta M_A\right)^2 + \left(\frac{\partial\Gamma_{2,i}}{\partial M_B}\right)^2 \left(\delta M_B\right)^2\nonumber\\ 
    &&+ \left(\frac{\partial\Gamma_{2,i}}{\partial M_C}\right)^2 \left(\delta M_C\right)^2,
\end{eqnarray}
where $\delta M_i$ is just the error of the mass of particle $i$ get from the previous subsection. Finally, the error of the total two-body decay width is calculated as
\begin{eqnarray}
    \delta\Gamma_{2} = \sqrt{\sum_i \left(\delta \Gamma_{2,i}\right)^2}.
\end{eqnarray}

\subsection{The possible competitions between one-gluon exchange and one-boson exchange}

In the early stage of this work, we tried to fit the ground baryon spectra with one-gluon-exchange only under SU(3) symmetry, and the model parameters and fit results are presented in Table~\ref{4-11} and Table~\ref{4-13}, respectively.

\begin{table}[!htb]
	\centering
	\caption{Quark model parameters used in the calculation, where the numbers with superscript "$\ast$" means they are previously fixed during the fit.\label{4-11}}
	\begin{tabular}{c|cc c}
		\hline \hline
	 
		&$m_{u,d}$ (MeV)   &~~~~313$^\ast$ & \\
		Quark masses    &$m_s$ (MeV)  &~~~~555$^\ast$ & \\
		&$m_c$ (MeV)  &~~~~1800$^\ast$ &\\ \hline
			Confinement &$a_{c}$ (MeV$\cdot$fm$^{-1}$)  &~~~~105.44$\pm$0.48469 & \\
			&$\Delta$ (MeV)  &~~~-79.501$\pm$0.67998 & 
		 \\ \hline
		&$\hat{r}_0=\hat{r}_g~$(MeV$\cdot$fm)  &~~~~30.633$\pm$0.21361 & \\
		&$\alpha_{uu}$  &~~~~0.878$\pm$0.0044 & \\
		&$\alpha_{us}$  &~~~~0.799$\pm$0.0039 & \\
		OGE &$\alpha_{ss}$  &~~~~0.758$\pm$0.0042 & \\
        &$\alpha_{uc}$ &~~~~0.750$\pm$0.0038 & \\
		&$\alpha_{sc}$  &~~~~0.691$\pm$0.0035 & \\ 
		&$\alpha_{cc}$  &~~~~0.666$\pm$0.0029 & \\
	  \hline \hline
	\end{tabular}
\end{table}

\begin{table}[!htb]
	\centering
	\caption{The masses of ground state baryons in unit of MeV.\label{4-13}}
	\begin{tabular}{ccccccccccccccccccc}
		\hline \hline
		~     &N~       &$\Delta$~  &$\Lambda$~     &$\Sigma$~          &$\Sigma^{*}$~      &$\Xi$~                                                  \\ \hline
		ChQM~ &942.12     &1227.97   &\textbf{1153.30}        &\textbf{1153.30}           &1386.69             &1328.21                                                                                                                            \\
		Expt~ &939    &1232      &1116~          &1189~             &1383~               &1318~                                                                        \\ \hline
		
		~    &$\Xi^{*}$  &$\Omega$~  &$\Lambda_c$~      &$\Sigma_{c}$~  &$\Sigma_{c}^{*}$~    &$\Xi_c$~    \\ \hline
		ChQM~  &1531.73  &1662.10~  &2276.85~    &2453.55~              &2514.81~         &2466.67~                 \\
		Expt~   &1533  &1672~  &2286~     &2455~             &2520~         &2471~                                                                                                         \\ \hline
		
		~  &$\Xi_c^{\prime}$~ &$\Xi_{c}^{\prime*}$~ &$\Omega_{c}$~  &$\Omega_{c}^{*}$                  \\ \hline
		ChQM~  &2582.66~       &2647.04~    &2696.18~      &2765.14~              \\
		Expt~   &2579~    &2645~    &2695~        &2770~  
         \\\hline
		\hline \hline
	\end{tabular}
\end{table}

It is easy to see that when considering the one-gluon-exchange only, the behavior of $\alpha_s$ does decrease with the momentum transfer, which obeys the requirement of quantum chromodynamics. However, at this time, $\Lambda$ and $\Sigma$ will have the same mass. Such a thing is actually understandable since the gluon exchange can not distinguish isospin of the hadron if we use SU(3) symmetry to describe $\Lambda$ and $\Sigma$, which makes us realize the necessity of the introduction of Goldstone exchange in baryon calculation if we want to adopt a unified SU(3) symmetry to describe both light and single heavy baryons. In addition, we also find that such a thing also happens in the meson sector. For example, $\rho$ and $\omega$ will also have the same mass when considering gluon exchange only, which shares the same reason as baryon fit.

Thus, we accept the chiral quark model and add into the Goldstone exchanges, and to obey the conclusion of chiral symmetry, we also fix the cutoffs previously that around 1 GeV. However, at this time, we find that fit on the ground baryon spectra will violate the usual behavior of $\alpha_s$. It is, in our view, because when the cutoffs of the Goldstone exchange is previously fixed, their contributions are also previously fixed, then, during the fit, the behavior of $\alpha_s$ is sacrificed by the program. In other words, there may exist some mysterious competitions that we have not figured out between gulon and Goldstone exchange contributions in light quarks interactions when fitting the ground baryon spectra. Thus, in the next step, we make all the cutoffs in the Goldstone exchanges free and fit the ground baryon spectra again, and find that although the behavior of $\alpha_s$ can be revealed, some of the cutoffs are smaller than the masses of the exchanged bosons, which we think is unacceptable since in our view it is not physical permitted. Finally, to obey the requirement of physics, we use the model parameters along with the fit result given in Table~\ref{4-0} and Table~\ref{4-3}.

\bibliographystyle{elsarticle-num}
\bibliography{ref}

\end{document}